\documentclass[journal=jctcce,manuscript=article]{achemso}
\usepackage[version=3]{mhchem}

\usepackage{amssymb,amsmath,amsfonts,multicol,multirow,longtable,array}
\usepackage{lscape}
 
\usepackage{appendix}
\usepackage{soul} 
\usepackage[usenames]{color}
\usepackage{makecell}
\usepackage{enumerate}
\usepackage{xr}
\usepackage[T1]{fontenc} 
\usepackage{booktabs}
\usepackage{tikz}
\usepackage{threeparttable}
\usepackage{graphicx}
\usepackage[figuresright]{rotating}
\usepackage{amsfonts}
\usepackage{colortbl}
\usepackage{framed}
\usepackage{verbatim}
\usepackage{amsbsy}
\usepackage{bm}
\usepackage{bbm}
\usepackage[normalem]{ulem}
\usepackage{float}
\usepackage[linesnumbered,boxed]{algorithm2e}
\usepackage{algorithm2e}
\usepackage{diagbox}
\usepackage{tabularx} 
\usepackage{booktabs} 
\usepackage{physics}
\usepackage{subcaption}

\usepackage[colorlinks=true, urlcolor=blue]{hyperref}
\usepackage{enumitem}
\setlist[enumerate]{label=(\arabic*), itemsep=2pt, parsep=0pt, topsep=3pt, partopsep=0pt}
\usepackage{xcolor}

\usepackage{braket}

\newcounter{xformulation}

\newcommand{\vct}[1]{\bm{#1}} 
\newcommand{\mat}[1]{\bm{#1}} 

\newfloat{Algorithm}{htbp}{alg}
\floatname{Algorithm}{Algorithm}

\newcounter{exe}[figure]
\newcommand{\iexe}{\refstepcounter{exe}\the\value{exe}:}

\setkeys{acs}{maxauthors = 0} 

\author{Xiaoyu Zhang} 
\email{zhangxiaoyu@stu.pku.edu.cn} 
\affiliation{College of Chemistry and Molecular Engineering, Peking University, Beijing 100871, the People's Republic of China} 

\title{End-to-End Differentiable Learning of a Single Functional for DFT and Linear-Response TDDFT}

\begin{document}

\begin{abstract}
 Density functional theory (DFT) and linear-response time-dependent density functional theory (LR‑TDDFT) rely on an exchange–correlation (xc) approximation that provides not only energy but also its functional derivatives that enter the self-consistent potential and the response kernel. Here, we present an end-to-end differentiable workflow to optimize a single deep-learned energy functional using targets from both Kohn–Sham DFT and adiabatic LR‑TDDFT. To enable this training in a computationally efficient and differentiable manner, we developed a JAX-based two-component quantum chemistry package (IQC), in which the learned functional provides a self-consistent potential and linear-response kernel via automatic differentiation. This construction permits gradient-based optimization through both the self-consistent-field (SCF) fixed-point equations and the Casida eigenvalue problem. We learn an exchange-correlation functional on excitation energies and ground-state properties (noncovalent interactions, thermochemistry, bond dissociation, ionization potentials, electron affinities, isomerization energies, and reaction barriers) while incorporating one-electron self-interaction cancelation as penalty terms, and we assess its possible transfer to molecular test cases. 
\end{abstract}

\section{Introduction}

Density functional theory (DFT) \cite{PhysRev.136.B864,KS-DFT} and its linear-response time-dependent extension (LR-TDDFT) \cite{PhysRevLett.52.997,tddft} provide an efficient route to ground-state and excitation energies for molecules and materials; yet, their predictive accuracy is ultimately limited by the exchange–correlation (xc) approximation. In Kohn–Sham DFT, the xc approximation enters the self-consistent field (SCF) equations through a first derivative (the potential contribution), while in adiabatic LR-TDDFT, it further enters through a second derivative (the response kernel) that governs excitation energies in the Casida formulation.

A central difficulty is that the same xc approximation must simultaneously control energies, SCF potentials, and LR kernels. Most traditional functionals are parameterized primarily against ground-state data, and their transferability to excited states is therefore not guaranteed. \cite{10.1063/1.1390175,10.1063/1.1904565,Mardirossian02102017,Zhao2008TheMS,YANAI200451}
A common workaround is to tune parameters (e.g., range-separation or hybrid mixing) for a specific system or class of systems to improve selected excitation energies, at the cost of reduced transferability. \cite{OKUNO201229,doi:10.1021/ct2009363} 
This limitation becomes even more acute for data-driven functionals \cite{10.1063/5.0150587,neuralxc,dm21,doi:10.1021/acs.jctc.0c00872,PhysRevLett.127.126403}, though previous works were done on the ground state: for example, DeepKS achieves substantially improved force accuracy only when forces are included explicitly as training targets, rather than emerging automatically from fitting energies alone. \cite{doi:10.1021/acs.jctc.0c00872} Motivated by this observation, we include LR excitation information explicitly in the training objective and optimize the functional end-to-end through both the SCF fixed point and the LR-TDDFT eigenvalue problem.

Recent efforts have begun to apply machine learning to time-dependent density functional theory; for example, by learning time-dependent exchange–correlation potentials from real-time densities in model systems. However, they focus on real-time TDDFT which has much heavier computation costs than LR-TDDFT. \cite{PhysRevA.101.050501} In this work, we pursue a complementary strategy: we learn an energy functional that is employed self-consistently for determining the ground state and that, via automatic differentiation, provides a consistent adiabatic linear-response (LR) kernel for computing excitation energies. A key aspect of our approach is that LR excitation information is incorporated explicitly into the training objective, rather than relying on a functional optimized solely with respect to ground-state observables. Furthermore, we impose exact constraints on self-interaction errors for exchange–correlation (xc) functionals through an appropriate penalty term. 

We first introduce a differentiable SCF+LR-TDDFT training framework and implement it in a JAX-based quantum chemistry package IQC, which preserves analytic consistency among the learned energy functional, its associated potential, and the resulting adiabatic kernel. We then employ this framework to train a deep-learning-based xc functional, termed IXC, using excitation energies and ground-state properties (noncovalent interactions, thermochemistry, bond dissociation, ionization potentials, electron affinities, isomerization energies, and reaction barriers), while simultaneously penalizing self-interaction errors. Finally, we assess the performance of the resulting functional on both excitation energies using established benchmark datasets, and self-interaction errors.

\section{Theory and Method}
\subsection{Differentiable DFT and LR-TDDFT}
The training of deep learning-based functionals for diverse target properties constitutes a non-trivial problem. A primary difficulty arises from the fact that conventional quantum chemistry packages are inherently non-differentiable. Consequently, it is challenging to derive analytical gradients for each computational step manually. The absence of such gradients precludes the straightforward application of gradient-based optimization methods for direct model training. Several tricks have been proposed to partially solve this problem. One approach is referred to as iterative training.\cite{neuralxc,doi:10.1021/acs.jctc.0c00872} Since the SCF component remains non-differentiable, the overall training protocol cannot be fully optimized via gradient-based methods. Consequently, the loss function may increase between successive iterations, and training on different target quantities often necessitates the use of distinct \textit{ad hoc} techniques. Another strategy involves performing the training directly on the xc potential \cite{doi:10.1021/acs.jpclett.9b02838}, which is obtained via the Wu–Yang inversion procedure \cite{10.1063/1.1535422}. In this framework, the model is trained using highly accurate reference electron densities. However, during the self-consistent field (SCF) iterations, the evolving density can deviate substantially from the training density, which may induce numerical instabilities and convergence difficulties in the SCF procedure. To enable robust and stable training, we implement a fully differentiable quantum chemistry package, IQC (intelligent quantum chemistry), which supports two-component DFT \cite{PhysRevResearch.5.013036,doi:10.1021/acs.jctc.5c01305} and two-component TDDFT \cite{RN69,doi:10.1021/acs.jctc.5c01272}. 'Two-component' refers to two-component wavefunctions. A key to making the program differentiable is implementing the two quantum chemistry methods in the framework of JAX \cite{jax}. Two pertinent software packages, JAXDFT \cite{PhysRevLett.126.036401} and DQC \cite{PhysRevLett.127.126403,10.1063/5.0076202}, have been introduced and used for training xc functionals in the literature. However, none of them currently implement the two methods considered in this work. JAXDFT and DQC are both limited to one-component DFT on ground states. 

A concise overview of the two-component formulations of DFT and LR-TDDFT is provided here. We use  $\Gamma, \Lambda, \Theta$ for two-component AO basis functions, $P, Q, R$ for two-component molecular orbitals, and  $I, J, K$ / $A,B,C$ for occupied / virtual orbitals. Throughout the theory section, scalar quantities are written in ordinary italic type, vectors in bold symbols, and matrices in bold symbols; for example, $\vct{\theta}$ and $\vct{X}_k$ denote vectors, whereas $\mat{F}$, $\mat{D}$, and $\mat{A}_{\vct{\theta}}$ denote matrices. Indexed quantities such as $F_{\Gamma\Lambda}$ and $D_{\Gamma\Lambda}$ denote scalar matrix elements.

The molecular orbitals are expressed in terms of two-component bases,
\begin{equation}
    P = C_{\Lambda P} \Lambda
\end{equation}
where $C_{\Lambda P}$ are the expansion coefficients of the orbital $P$ in the basis $\Lambda$.

The first-order reduced density matrix is defined as follows:
\begin{equation}
   D_{\Lambda \Gamma} =  C_{\Lambda I}^{*} C_{\Gamma I}
\end{equation}

The matrix representation of the Fock operator for two-component DFT is given by:
\begin{equation}
\begin{split}
    F_{\Gamma \Lambda}& = h_{\Gamma \Lambda} + (\Gamma \Lambda | \Pi \Theta)D_{\Theta \Pi} -c_{\mathrm{HF}}
    (\Gamma \Theta | \Pi \Lambda)D_{\Theta \Pi} 
   + \frac{\partial E_{\mathrm{xc}}}{\partial D_{\Gamma \Lambda}}
    \end{split}
    \label{eq:fock}
\end{equation}
where $c_{\mathrm{HF}}$ is the percentage of the Hartree-Fock exchange in a hybrid functional.
By applying partial differentiation to the Fock matrix with respect to the density matrix, we obtain the matrix representation of the kernel operator:
\begin{equation}
        K_{\Gamma \Lambda \Theta \Pi}= (\Gamma \Lambda | \Pi \Theta)-c_{\mathrm{HF}} (\Gamma \Theta|\Pi \Lambda) + \frac{\partial^2 E_{\mathrm{xc}}}{\partial D_{\Gamma \Lambda} \partial D_{\Pi \Theta}}
        \label{eq:kernel}
\end{equation}

The two-component KS-DFT equation reads as follows:
\begin{equation}
    \mat{F}\mat{C}= \mat{S} \mat{C} \mat{\epsilon}
    \label{eq:scf}
\end{equation}
Here, $\mat{S}$ is the overlap matrix in a set of two-component basis functions. $\mat{\epsilon}$ is a diagonal matrix representing the orbital energies. By solving this equation within the SCF framework, we obtain the ground-state wavefunctions. Furthermore, to perform TDDFT calculations, it is necessary to first carry out a DFT calculation to determine the wavefunctions of the reference state.

The Casida equation for TDDFT reads as follows:
\begin{equation}
    \begin{pmatrix}
        H_{AIBJ} & H_{AIJB}\\
        H_{IABJ} & H_{IAJB}
    \end{pmatrix}
    \begin{pmatrix}
        X_{BJ}\\
        X_{JB}
    \end{pmatrix}
    =
    \begin{pmatrix}
        1 & 0\\
        0 & -1
    \end{pmatrix}
    \begin{pmatrix}
        X_{AI}\\
        X_{IA}
    \end{pmatrix}
    \Omega
\end{equation}
where 
\begin{equation}
\begin{split}
    \begin{pmatrix}
        H_{AIBJ} & H_{AIJB}\\
        H_{IABJ} & H_{IAJB}
    \end{pmatrix}
   & =
    \begin{pmatrix}
        F_{AB} \delta_{IJ} -F_{JI}\delta_{AB} & 0\\
        0 & F_{BA} \delta_{IJ}-F_{IJ}\delta_{AB}
    \end{pmatrix}
   \\ &+
    \begin{pmatrix}
        K_{AIBJ}& K_{AIJB}\\
        K_{IABJ}& K_{IAJB}
    \end{pmatrix}
    \end{split}
\end{equation}

In this work, we set $H_{AIJB}$ and $H_{IABJ}$ to be zero during training, which is referred to as the Tamm-Dancoff approximation (TDA). \cite{HIRATA1999291} Here, the fock and kernel are both under MO representation, which can be easily transformed from AO representation. In this work, we focus on adiabatic LR‑TDDFT, where the response kernel is obtained as the second derivative of the same energy functional used in the ground-state calculation. The frequency dependence (memory effects) of the exact TDDFT kernel is beyond the present scope; \cite{RevModPhys.74.601} our goal here is to enable joint optimization of ground-state and LR excitation targets while maintaining analytic consistency between the learned energy, potential, and adiabatic kernel.

There are two technical problems in backward differentiation. The first is the
treatment of the SCF loop. Prior approaches have employed a fixed number of
linear-mixing iterations to enforce self-consistency and then backpropagated
through the entire unrolled sequence. \cite{doi:10.1021/acscentsci.7b00586,PhysRevLett.126.036401}
However, straightforward unrolling may generate ill-defined intermediate states
and lead to memory consumption that grows with the number of SCF iterations,
which limits such approaches to very small systems such as $\mathrm{H}_2$. We
instead treat the converged SCF solution as the root of a fixed-point equation
in Fock space. The self-consistent solution
$\mat{F}_{\mathrm{scf}}(\vct{\theta})$ satisfies
\begin{equation}
g(\mat{F},\vct{\theta})\equiv \mathcal{S}_{\vct{\theta}}(\mat{F})-\mat{F}=B_{\vct{\theta}}(\mat{D}(\mat{F}))-\mat{F}=0.
\label{eq:scf_root}
\end{equation}
Here, $\mathcal{S}_{\vct{\theta}}$ denotes one SCF update that maps an
input Hermitian Fock matrix $\mat{F}$ to an output Fock matrix
$\mat{F}^{\mathrm{out}}$ and $B_{\vct{\theta}}$ denotes the Fock-building map. Note that $\mathcal{S}_{\vct{\theta}}$ does
not include mixing or DIIS/Anderson acceleration.

In the implementation, one application of $\mathcal{S}_{\vct{\theta}}$ proceeds as follows.
\begin{enumerate}
\item Choose a fixed orthogonalizer $\mat{O}$ satisfying
\begin{equation}
\mat{O}^\dagger \mat{S} \mat{O} = \mat{I}.
\label{eq:appA_orth}
\end{equation}

\item Transform the generalized KS equation to the ordinary Hermitian eigenvalue problem
\begin{equation}
\widetilde{\mat{F}} \mat{U} = \mat{U} \mat{\epsilon},\qquad \widetilde{\mat{F}}=\mat{O}^\dagger \mat{F} \mat{O}.
\label{eq:appA_stdks}
\end{equation}

\item Construct the density matrix from the eigensolution. In the convention of the density matrix
used throughout this work,
\begin{equation}
\mat{D}(\mat{F})=\mat{C}^* \mat{N} \mat{C}^T,\qquad \mat{C}=\mat{O}\mat{U},
\label{eq:appA_dm_from_eig}
\end{equation}
equivalently,
\begin{equation}
\mat{D}(\mat{F})=\mat{O}^* \mat{U}^* \mat{N} \mat{U}^T \mat{O}^T,
\end{equation}
where $\mat{N}$ is the diagonal occupation matrix.

\item Rebuild the Fock matrix from this density,
\begin{equation}
\mat{F}^{\mathrm{out}}=B_{\vct{\theta}}(\mat{D}(\mat{F})).
\label{eq:appA_buildfock}
\end{equation}
\end{enumerate}

Gradients of any scalar objective that depend on the SCF solution are then
obtained by implicit differentiation of Eq.~(\ref{eq:scf_root}), which
reduces the backward pass to the solution of an adjoint linear system involving
the Jacobian of $g$. Only Jacobian–vector and vector–Jacobian products of the
one-step map $\mathcal{S}_{\vct{\theta}}$ are required, so the memory cost is
independent of the number of forward SCF iterations.

The second issue is the differentiation of the eigendecomposition itself in the
presence of (near-)degenerate eigenvalues, which is common in solving both the
KS equation. Because JAX does not provide a built-in Jacobian-vector product (JVP) rule for the generalized eigendecomposition in Eq.~(\ref{eq:scf}), we do
not differentiate that generalized problem directly. Instead, after the
orthogonalization step above, we differentiate the resulting ordinary Hermitian
eigendecomposition with a custom JVP based on first-order perturbation theory.
The main difficulty is to regularize terms of the form
$\frac{1}{\Delta_{PQ}}$, where $\Delta_{PQ}$ denotes the gap between two
eigenvalues $P$ and $Q$. \cite{https://doi.org/10.1002/qua.560160825} In the
implementation, when $\Delta_{PQ}$ is very small, we replace
$\frac{1}{\Delta_{PQ}}$ by
$\frac{\Delta_{PQ}}{\Delta_{PQ}^2+\epsilon}$, where
$\epsilon$ is a small perturbative parameter set to $10^{-12}$ in this work.
Full derivations and numerical details on the two issues are provided in the next subsection.

\subsection{Implicit Differentiation through the SCF Fixed Point}

In this subsection, we present the backward differentiation of SCF fixed point and its implementation within the IQC framework. Throughout this subsection, the nuclear geometry,
the AO basis, and the overlap matrix are fixed, and differentiation is taken
only with respect to the functional parameters $\vct{\theta}$.

Let $\mathbb{H}_n=\{\mat{F}\in\mathbb{C}^{n\times n}:\mat{F}=\mat{F}^\dagger\}$ denote the space of
Hermitian Fock matrices. We equip this space with the Frobenius inner product
\begin{equation}
\langle \mat{A},\mat{B}\rangle \equiv \mathrm{Re}\,\mathrm{Tr}(\mat{A}^\dagger \mat{B}).
\end{equation}

Now let
\begin{equation}
\mathcal{L}(\vct{\theta})=\widetilde{\mathcal{L}}(\mat{F}_{\mathrm{scf}}(\vct{\theta}),\vct{\theta})
\label{eq:appA_loss}
\end{equation}
be any real-valued scalar objective depending on the converged SCF solution. Assuming that an infinitesimal perturbation $d\vct{\theta}$ is applied to $\vct{\theta}$, the parameters are modified to $\vct{\theta} + d\vct{\theta}$, which in turn yields a newly converged Fock matrix $\mat{F}_{\mathrm{scf}} + d\mat{F}_{\mathrm{scf}}$. By expanding the condition $g(\mat{F}_{\mathrm{scf}} + d\mat{F}_{\mathrm{scf}}, \vct{\theta} + d\vct{\theta}) = 0$ to first order in the perturbations, we obtain
\begin{equation}
  g(\mat{F}_{\mathrm{scf}}+d \mat{F}_{\mathrm{scf}},\vct{\theta}+d\vct{\theta})=  g(\mat{F}_{\mathrm{scf}},\vct{\theta}) + \frac{\partial g}{\partial \mat{F}}|_{\mat{F}_{\mathrm{scf}}}
  [d \mat{F}_{\mathrm{scf}}] + \frac{\partial g}{\partial \vct{\theta}}|_{\mat{F}_{\mathrm{scf}}} [d \vct{\theta}] + \mathcal{O}(||d||^2).
\end{equation}

Differentiating Eq.~(\ref{eq:scf_root}) gives
\begin{equation}
\left(
\mat{I}-\frac{\partial \mathcal{S}_{\vct{\theta}}}{\partial \mat{F}}\bigg|_{\mat{F}_{\mathrm{scf}}}
\right)d\mat{F}_{\mathrm{scf}}
=
\frac{\partial \mathcal{S}_{\vct{\theta}}}{\partial \vct{\theta}}\bigg|_{\mat{F}_{\mathrm{scf}}} d\vct{\theta}.
\label{eq:appA_tangent_eq}
\end{equation}
It is convenient to define the linear operator
\begin{equation}
\mathcal{A}
\equiv
\mathcal{I}
-
\left.
\partial_{\mat{F}}\mathcal{S}_{\vct{\theta}}
\right|_{\mat{F}_{\mathrm{scf}}},
\label{eq:appA_Adef}
\end{equation}
where $\mathcal{I}$ is the identity operator on $\mathbb{H}_n$ and
$\partial_{\mat{F}}\mathcal{S}_{\vct{\theta}}$ denotes the Fr\'echet derivative of
the one-step SCF map with respect to the input Fock matrix. Equation~(\ref{eq:appA_tangent_eq})
is then
\begin{equation}
\mathcal{A}\!\left[d\mat{F}_{\mathrm{scf}}\right]
=
\left.
\partial_{\vct{\theta}}\mathcal{S}_{\vct{\theta}}
\right|_{\mat{F}_{\mathrm{scf}}}
\!\left[d\vct{\theta}\right].
\label{eq:as}
\end{equation}

Directly forming or inverting $\mathcal{A}$ is unnecessary. In reverse-mode
differentiation, we instead introduce an adjoint matrix
$\mat{\lambda}\in\mathbb{H}_n$ satisfying
\begin{equation}
\mathcal{A}^{*}\!\left[\mat{\lambda}\right]
=
\nabla_{\mat{F}}
\widetilde{\mathcal{L}}(\mat{F}_{\mathrm{scf}},\vct{\theta}).
\label{eq:appA_adjoint_eq}
\end{equation}
Here $\mathcal{A}^{*}$ is the adjoint of the linear operator $\mathcal{A}$ with
respect to the Frobenius inner product
$\langle \mat{X},\mat{Y}\rangle=\operatorname{Re}\operatorname{Tr}(\mat{X}^{\dagger}\mat{Y})$.

By application of the chain rule, the total derivative of \(\mathcal{L}\) with respect to the model parameters \(\vct{\theta}\) is given by
\begin{equation}
d\mathcal{L}\!\left[d\vct{\theta}\right]
=
\left.
\partial_{\vct{\theta}}\widetilde{\mathcal{L}}
\right|_{\mat{F}_{\mathrm{scf}}}
\!\left[d\vct{\theta}\right]
+
\left\langle
\nabla_{\mat{F}} \widetilde{\mathcal{L}} ,\,
d\mat{F}_{\mathrm{scf}}
\right\rangle .
\end{equation}
The explicit occurrence of \(d \mat{F}_{\mathrm{scf}}\) can be eliminated by observing that
\begin{equation}
    \left\langle
\nabla_{\mat{F}} \widetilde{\mathcal{L}} ,\,
d\mat{F}_{\mathrm{scf}}
\right\rangle
=
\left\langle
\vphantom{\nabla_{\mat{F}} \widetilde{\mathcal{L}}}\mathcal{A}^* [\lambda] ,\,
d\mat{F}_{\mathrm{scf}}
\right\rangle
=
\left\langle
\vphantom{\nabla_{\mat{F}} \widetilde{\mathcal{L}}}\lambda ,\,
\mathcal{A}[d\mat{F}_{\mathrm{scf}}]
\right\rangle
=
\left\langle
\vphantom{\nabla_{\mat{F}} \widetilde{\mathcal{L}}}\lambda ,\,
\left.
\partial_{\vct{\theta}}\mathcal{S}_{\vct{\theta}}
\right|_{\mat{F}_{\mathrm{scf}}}
\!\left[d\vct{\theta}\right]
\right\rangle ,
\end{equation}
The validity of the first step follows from Eq. (\ref{eq:appA_adjoint_eq}). The justification of the second step is provided by the definition of the adjoint operator. The correctness of the third step is ensured by Eq. (\ref{eq:as}). Finally, we have the total derivative with respect to the model parameters
\begin{equation}
d\mathcal{L}\!\left[d\vct{\theta}\right]
=
\left.
\partial_{\vct{\theta}}\widetilde{\mathcal{L}}
\right|_{\mat{F}_{\mathrm{scf}}}
\!\left[d\vct{\theta}\right]
+
\left\langle
\mat{\lambda},\,
\left.
\partial_{\vct{\theta}}\mathcal{S}_{\vct{\theta}}
\right|_{\mat{F}_{\mathrm{scf}}}
\!\left[d\vct{\theta}\right]
\right\rangle .
\label{eq:appA_total_diff_A}
\end{equation}
Equivalently, in vector-gradient notation,
\begin{equation}
\nabla_{\vct{\theta}}\mathcal{L}
=
\nabla_{\vct{\theta}}\widetilde{\mathcal{L}}
+
\left(
\left.
\partial_{\vct{\theta}}\mathcal{S}_{\vct{\theta}}
\right|_{\mat{F}_{\mathrm{scf}}}
\right)^{*}
\!\left[\mat{\lambda}\right].
\label{eq:appA_total_grad_A}
\end{equation}

This is equivalent to the formulation written directly in terms of
$g(\mat{F},\vct{\theta})$, since
\begin{equation}
\left.
\partial_{\mat{F}}g
\right|_{\mat{F}_{\mathrm{scf}}}
=
\left.
\partial_{\mat{F}}\mathcal{S}_{\vct{\theta}}
\right|_{\mat{F}_{\mathrm{scf}}}
-
\mathcal{I}
=
-\mathcal{A},
\qquad
\left.
\partial_{\vct{\theta}}g
\right|_{\mat{F}_{\mathrm{scf}}}
=
\left.
\partial_{\vct{\theta}}\mathcal{S}_{\vct{\theta}}
\right|_{\mat{F}_{\mathrm{scf}}}.
\label{eq:appA_gA_relation}
\end{equation}
Hence, if one defines $\mat{\Lambda}=-\mat{\lambda}$, then
\begin{equation}
\left(
\left.
\partial_{\mat{F}}g
\right|_{\mat{F}_{\mathrm{scf}}}
\right)^{*}
\!\left[\mat{\Lambda}\right]
=
\nabla_{\mat{F}}
\widetilde{\mathcal{L}}(\mat{F}_{\mathrm{scf}},\vct{\theta})
\label{eq:appA_adjoint_eq_g}
\end{equation}
and
\begin{equation}
d\mathcal{L}\!\left[d\vct{\theta}\right]
=
\left.
\partial_{\vct{\theta}}\widetilde{\mathcal{L}}
\right|_{\mat{F}_{\mathrm{scf}}}
\!\left[d\vct{\theta}\right]
-
\left\langle
\mat{\Lambda},\,
\left.
\partial_{\vct{\theta}}g
\right|_{\mat{F}_{\mathrm{scf}}}
\!\left[d\vct{\theta}\right]
\right\rangle .
\label{eq:appA_total_diff_g}
\end{equation}
Equivalently,
\begin{equation}
\nabla_{\vct{\theta}}\mathcal{L}
=
\nabla_{\vct{\theta}}\widetilde{\mathcal{L}}
-
\left(
\left.
\partial_{\vct{\theta}}g
\right|_{\mat{F}_{\mathrm{scf}}}
\right)^{*}
\!\left[\mat{\Lambda}\right].
\label{eq:appA_total_grad_g}
\end{equation}

In practice, we never construct the Jacobian matrices explicitly. For any matrix
perturbation $\mat{V}\in\mathbb{H}_n$, the action of the Jacobian is evaluated as a
directional derivative,
\begin{equation}
\left.
\partial_{\mat{F}}\mathcal{S}_{\vct{\theta}}
\right|_{\mat{F}_{\mathrm{scf}}}
\!\left[\mat{V}\right]
=
\left.
\frac{d}{d\eta}\,
\mathcal{S}_{\vct{\theta}}(\mat{F}_{\mathrm{scf}}+\eta\mat{V})
\right|_{\eta=0},
\label{eq:appA_step_jvp}
\end{equation}
which is obtained by automatic differentiation of the single-step map
$\mathcal{S}_{\vct{\theta}}$, with the eigendecomposition differentiated by the custom
rule. Derivations are presented as follows.

Although the SCF problem is naturally a generalized eigenvalue problem,
the actual differentiation in the code is performed after reducing it to the
ordinary Hermitian problem in Eq.~(\ref{eq:appA_stdks}). In particular, we do
not rely on a built-in JVP rule for a generalized eigendecomposition.

At zero temperature and away from occupied-virtual crossings, the occupation
pattern is locally constant, so $\mat{N}$ can be treated as fixed in the
differentiation. Then the derivative of $\mat{D}(\mat{F})$ is determined by the derivative
of the eigendecomposition in Eq.~(\ref{eq:appA_stdks}). For a perturbation
$\delta \widetilde{\mat{F}}$, let
\begin{equation}
\mat{M}=\mat{U}^\dagger (\delta \widetilde{\mat{F}}) \mat{U}.
\label{eq:appA_Mdef}
\end{equation}
In first-order perturbation theory,
\begin{equation}
\delta \epsilon_P = \mathrm{Re}\, M_{PP},
\label{eq:appA_eval_jvp}
\end{equation}
and the eigenvector response is
\begin{equation}
\delta \mat{U} = \mat{U} (\mat{R} \odot \mat{M}),
\label{eq:appA_evec_jvp}
\end{equation}
where $\odot$ denotes the Hadamard product $[(\mat{A}\odot \mat{B})_{ij}=A_{ij}B_{ij}]$ and
\begin{equation}
R_{PQ}=
\begin{cases}
\dfrac{1}{\epsilon_Q-\epsilon_P}, & P\neq Q,\\[6pt]
0, & P=Q.
\end{cases}
\end{equation}
When the eigenvalue gap becomes very small, the factor
$1/(\epsilon_Q-\epsilon_P)$ is regularized in the implementation. Specifically,
for $\Delta_{PQ}=\epsilon_Q-\epsilon_P$ we use
\begin{equation}
R_{PQ}=
\begin{cases}
\dfrac{1}{\Delta_{PQ}}, & |\Delta_{PQ}|\ge \tau,\\[10pt]
\dfrac{\Delta_{PQ}}{\Delta_{PQ}^2+\varepsilon}, & |\Delta_{PQ}|<\tau,
\end{cases}
\qquad \tau=10^{-5},\qquad \varepsilon=10^{-12}.
\label{eq:appA_gap_regularization}
\end{equation}
This is the custom JVP rule used for the Hermitian eigensolver. Therefore, the
JVP of the one-step SCF map $\mathcal{S}_{\vct{\theta}}$ is obtained by automatic
differentiation of Eqs.~(\ref{eq:appA_dm_from_eig}) and
(\ref{eq:appA_buildfock}), using the regularized eigendecomposition rule above.

Under the fixed-occupation assumption, the density response can be written as
\begin{equation}
\delta \mat{D}
=
\mat{O}^*\left(
\delta \mat{U}^*\, \mat{N} \mat{U}^T + \mat{U}^* \mat{N}\, \delta \mat{U}^T
\right)\mat{O}^T.
\label{eq:appA_dm_jvp}
\end{equation}

Equivalently, one can regard this as the directional derivative of the map
$\mat{F}\mapsto \mat{D}(\mat{F})$ and hence of the full one-step map
$\mat{F}\mapsto \mathcal{S}_{\vct{\theta}}(\mat{F})$.

\subsection{Fixed-Density Backpropagation for DFT Total Energies}
\label{subsec:fixed_density_energy_backward}

The implicit fixed-point differentiation described above is the general
backward route in IQC: any scalar objective that depends on the converged SCF
solution can be differentiated by solving the SCF adjoint equation.  For
objectives that depend only on converged total energies, IQC also provides a
less expensive fixed-density backward route.  This route is not an additional
physical approximation to the forward calculation; the forward SCF is still
solved self-consistently.  The simplification is only in the reverse pass and
follows from the variational stationarity of the SCF total energy.

In the present implementation, we set $c_{\mathrm{HF}}=1$ and learn only an additional interaction correction beyond full Hartree–Fock exchange. To distinguish it from a conventional semilocal xc term, we denote the learned scalar energy by
$E_{\mathrm{IXC}}(\mat{D};\vct{\theta})$, where $\mat{D}$ is the one-particle density matrix and $\vct{\theta}$ collects all network parameters. The total energy used in the SCF procedure is therefore
\begin{equation}
E(\mat{D};\vct{\theta})=E_{1\mathrm{b}}[\mat{D}]+E_{J}[\mat{D}]-E_{K}[\mat{D}]+E_{\mathrm{IXC}}(\mat{D};\vct{\theta})+E_{\mathrm{nuc}}.
\label{eq:ixc_total_energy}
\end{equation}

The self-consistent density \(\mat{D}_{\mathrm{scf}}(\vct{\theta})\) is a stationary point of
\(E(\mat{D}(\mat{C});\vct{\theta})\) on the manifold of \(\mat{S}\)-orthonormal orbitals with the fixed
occupation pattern, where $\mat{C}$ is the coefficient matrix of the molecular orbitals.  Equivalently, for an occupied--virtual orbital rotation matrix \(\mat{\kappa}\), with elements \(\kappa_{AI}\),
\begin{equation}
\left.
\frac{\partial E(\mat{D}(\mat{C}(\mat{\kappa}));\vct{\theta})}{\partial \kappa_{AI}}
\right|_{\mat{\kappa}=\mat{0}}
=
0 .
\label{eq:energy_stationarity}
\end{equation}
Using
\begin{equation}
[\mat{F}(\mat{D};\vct{\theta})]_{\Gamma\Lambda}
=
\frac{\partial E(\mat{D};\vct{\theta})}{\partial D_{\Gamma\Lambda}},
\label{eq:fixed_density_fock_as_energy_derivative}
\end{equation}
the first-order energy variation is
\begin{equation}
\delta E(\mat{D};\vct{\theta})
=
\left\langle \mat{F}(\mat{D};\vct{\theta}),\delta \mat{D}\right\rangle
\label{eq:fixed_density_first_variation}
\end{equation}
Equation~(\ref{eq:energy_stationarity}) therefore implies
\begin{equation}
\left.
\left\langle \mat{F},\delta \mat{D}\right\rangle
\right|_{\mat{D}=\mat{D}_{\mathrm{scf}}(\vct{\theta})}
=0
\label{eq:fixed_density_stationary_inner_product}
\end{equation}
for every first-order density variation \(\delta \mat{D}\) induced by an allowed
orbital rotation.

Now consider the converged total energy
\begin{equation}
E_{\mathrm{scf}}(\vct{\theta})
=
E(\mat{D}_{\mathrm{scf}}(\vct{\theta});\vct{\theta})
\end{equation}
Differentiating it with respect to the model parameters gives
\begin{equation}
\frac{dE_{\mathrm{scf}}(\vct{\theta})}{d\vct{\theta}}
=
\left.
\frac{\partial E(\mat{D};\vct{\theta})}{\partial \vct{\theta}}
\right|_{\mat{D}=\mat{D}_{\mathrm{scf}}}
+
\left.
\left\langle
\frac{\partial E (\mat{D};\vct{\theta})}{\partial \mat{D}},
\frac{d\mat{D}_{\mathrm{scf}}(\vct{\theta})}{d\vct{\theta}}
\right\rangle
\right|_{\mat{D}=\mat{D}_{\mathrm{scf}}}\ .
\label{eq:fixed_density_chain_rule}
\end{equation}
Then, the working equation for total-energy backpropagation is
\begin{equation}
\frac{dE_{\mathrm{scf}}(\vct{\theta})}{d\vct{\theta}}
=
\left.
\frac{\partial E_{\mathrm{IXC}}(\mat{D};\vct{\theta})}{\partial\vct{\theta}}
\right|_{\mat{D}=\mat{D}_{\mathrm{scf}}} .
\label{eq:fixed_density_energy_gradient}
\end{equation}

The implementation realizes Eq.~(\ref{eq:fixed_density_energy_gradient}) with a
surrogate scalar whose value is the converged SCF energy but whose derivative is
evaluated at fixed density.  

This is the route used for total-energy penalties and it requires a sufficiently converged SCF solution and an unchanged
occupation pattern.  It is not used for excitation-energy targets, because an
excitation energy is an eigenvalue of a response operator rather than a
stationary ground-state total energy.

\subsection{Backpropagation through TDA Excitation Energies}
\label{subsec:tda_excitation_backward}

Excitation-energy objectives require a different backward equation.  The
quantity being fitted is not the stationary ground-state energy but an
eigenvalue of the TDA equation constructed from the converged SCF
reference.  Consequently, its derivative contains the derivative of the
response operator itself, together with the implicit derivative of the
self-consistent reference density from the previous subsection.

The TDA excitation energies are the eigenvalues
\begin{equation}
 \mat{A}_{\vct{\theta}} \vct{X}_k
=
\Omega_k \vct{X}_k .
\label{eq:tda_eigenproblem_iqc}
\end{equation}
In the forward pass, IQC obtains \(\Omega_k\) and \(\vct{X}_k\) with an iterative
Davidson solver.  The Davidson iterations are not unrolled in reverse mode.
Instead, once a root has been found, the backward pass uses the stationary
Rayleigh quotient with the response vector held fixed:
\begin{equation}
R_k(\vct{\theta})
=
\frac{
\langle \vct{X}_k, \mat{A}_{\vct{\theta}} \vct{X}_k\rangle
}{
\langle \vct{X}_k,\vct{X}_k\rangle
},
\label{eq:tda_rayleigh_iqc}
\end{equation}
where \(\vct{X}_k\) is treated as a stop-gradient quantity.  For an isolated root,
the Hellmann–Feynman derivative is
\begin{equation}
d\Omega_k
=
\frac{
\left\langle
\vct{X}_k,
\left(d \mat{A}_{\vct{\theta}}\right)\vct{X}_k
\right\rangle
}{
\langle \vct{X}_k,\vct{X}_k\rangle
}.
\label{eq:tda_hf_derivative}
\end{equation}

For a cluster of nearly degenerate roots \(\mathcal G\), the individual
Davidson vectors are not unique.  We therefore replace the single-vector
Rayleigh quotient by the trace quotient over the selected response subspace:
\begin{equation}
R_{\mathcal G}(\vct{\theta})
=
\frac{1}{|\mathcal G|}
\operatorname{Tr}
\left[
\left(\mat{Q}_{\mathcal G}^{\dagger}\mat{Q}_{\mathcal G}\right)^{+}
\mat{Q}_{\mathcal G}^{\dagger}
\mat{A}_{\vct{\theta}}
\mat{Q}_{\mathcal G}
\right],
\label{eq:tda_degenerate_trace_iqc}
\end{equation}
where \(\mat{Q}_{\mathcal G}\) contains the response vectors in the cluster and
\((\cdot)^+\) denotes the Moore--Penrose pseudoinverse.  This expression is
invariant to unitary rotations within the degenerate subspace.

Combining the TDA derivative with the SCF fixed-point derivative gives the
parameter derivative of an excitation energy:
\begin{equation}
\frac{d\Omega_k}{d\vct{\theta}}
=
\left\langle
\frac{\partial \Omega_k}{\partial  \mat{A}_{\vct{\theta}}},
\frac{\partial  \mat{A}_{\vct{\theta}}}{\partial\vct{\theta}}
+
\frac{\partial  \mat{A}_{\vct{\theta}}}{\partial \mat{D}_{\mathrm{scf}}}
\frac{d\mat{D}_{\mathrm{scf}}}{d\vct{\theta}}
\right\rangle .
\label{eq:tda_total_parameter_derivative}
\end{equation}
The density-response term
\(d\mat{D}_{\mathrm{scf}}/d\vct{\theta}\) is supplied by the implicit SCF fixed-point equation from
the previous subsection.  During this TDA backward pass, the occupied
–virtual response basis and the selected response vectors are held fixed; this fixes the MO gauge and avoids derivatives of arbitrary rotations inside nearly degenerate occupied, virtual, or response subspaces. 

\subsection{Modeling}

The expression for the total energy is given in Eq.~(\ref{eq:ixc_total_energy}). The learned
contribution to the self-consistent potential and to the adiabatic LR kernel is obtained from
the first and second derivatives of the same scalar functional,
\begin{equation}
F^{\mathrm{IXC}}_{\Gamma\Lambda}(\mat{D};\vct{\theta})
=
\frac{\partial E_{\mathrm{IXC}}(\mat{D};\vct{\theta})}
{\partial D_{\Gamma\Lambda}},
\qquad
K^{\mathrm{IXC}}_{\Gamma\Lambda\Theta\Pi}(\mat{D};\vct{\theta})
=
\frac{\partial F^{\mathrm{IXC}}_{\Gamma\Lambda}}
{\partial D_{\Pi\Theta}}
=
\frac{\partial^2 E_{\mathrm{IXC}}(\mat{D};\vct{\theta})}
{\partial D_{\Gamma\Lambda}\partial D_{\Pi\Theta}} .
\label{eq:ixc_derivatives}
\end{equation}
These quantities are inserted into Eqs.~(\ref{eq:fock}) and (\ref{eq:kernel}). Thus, the
model learns a single energy functional rather than fitting the potential and the response
kernel as independent objects.

The model takes as input the density matrix at each SCF step. For a two-component
calculation, the spinor density is first written in block form,
\begin{equation}
\mat{D}=
\begin{pmatrix}
\mat{D}^{\alpha\alpha} & \mat{D}^{\alpha\beta}\\
\mat{D}^{\beta\alpha} & \mat{D}^{\beta\beta}
\end{pmatrix},
\label{eq:ixc_spinor_dm}
\end{equation}
and decomposed into one charge channel and three spin-magnetization channels,
\begin{equation}
\mat{D}^{(0)}
=
\mat{D}^{\alpha\alpha}+\mat{D}^{\beta\beta},
\quad
\mat{D}^{(x)}
=
\mat{D}^{\alpha\beta}+\mat{D}^{\beta\alpha}, \quad
\mat{D}^{(y)}
=
i\left(-\mat{D}^{\alpha\beta}+\mat{D}^{\beta\alpha}\right),
\quad
\mat{D}^{(z)}
=
\mat{D}^{\alpha\alpha}-\mat{D}^{\beta\beta}.
\label{eq:ixc_spin_components}
\end{equation}
This decomposition gives the model an input with direct physical meaning: the network
does not see arbitrary spinor matrix elements, but charge and spin densities represented in
an auxiliary basis.

To construct this representation, a fixed auxiliary basis $\{\chi_M\}$ is introduced and the
density is projected onto it using three-center Coulomb integrals. In all calculations below,
the descriptor basis is the Weigend auxiliary basis \cite{B515623H}. This auxiliary basis is
used only to construct descriptors for $E_{\mathrm{IXC}}$; it is not used as a density-fitting
approximation for the Coulomb or exchange terms in the SCF solver. Let
\begin{equation}
B_{M,\mu\nu}
=
(M|\mu\nu)
=
\iint
\frac{\chi_M(\mathbf r_1)\mu(\mathbf r_2)\nu(\mathbf r_2)}
{r_{12}}
\,d\mathbf r_1\,d\mathbf r_2,
\qquad
V_{MN}=(M|N),
\label{eq:ixc_three_center}
\end{equation}
where $\mu,\nu$ denote spatial AO indices and $\mat{V}$ is the Coulomb metric in the
auxiliary space. The required integral tensors are evaluated with PySCF
\cite{10.1063/5.0006074}.

The auxiliary functions are grouped by atom $a$ and angular momentum $\ell$. Within each
$(a,\ell)$ block, only the radial channels are orthogonalized, while the angular channels are
kept explicit. Writing the auxiliary index as $(a,\ell,r,m)$, where $r$ labels the radial shell
and $m=-\ell,\ldots,\ell$ labels the magnetic component, the radial block metric is
\begin{equation}
W^{(a\ell)}_{rs}
=
\frac{1}{2\ell+1}
\sum_{m=-\ell}^{\ell}
V_{(a\ell r m),(a\ell s m)} .
\label{eq:ixc_radial_metric}
\end{equation}

A numerical truncation is done here. Let
\begin{equation}
\mat{W}^{(a\ell)}
=
\mat{U}^{(a\ell)}
\operatorname{diag}\!\left(w_i^{(a\ell)}\right)
[\mat{U}^{(a\ell)}]^T
\end{equation}
be its eigenvalue decomposition. In the whitening step, we do not use the
bare inverse square root. Instead, we define
\begin{equation}
\left[\mat{W_+}^{(a\ell)}\right]^{-1/2}
=
\mat{U}^{(a\ell)}
\operatorname{diag}
\left(
\begin{cases}
\left(w_i^{(a\ell)}\right)^{-1/2},
&
w_i^{(a\ell)}>\tau_W^{(a\ell)},\\[4pt]
0,
&
w_i^{(a\ell)}\le \tau_W^{(a\ell)}
\end{cases}
\right)
[\mat{U}^{(a\ell)}]^T,
\label{eq:radial_cut_inv_sqrt}
\end{equation}
with
\begin{equation}
\tau_W^{(a\ell)}
=
10^{-10}
\max\!\left(\max_i |w_i^{(a\ell)}|,1\right).
\label{eq:radial_cutoff}
\end{equation}

The three-center tensor is whitened only in this radial space,
\begin{equation}
\widetilde B_{(a\ell r m),\mu\nu}
=
\sum_s
\left[\mat{W_+}^{(a\ell)}\right]^{-1/2}_{rs}
B_{(a\ell s m),\mu\nu},
\label{eq:ixc_radial_whiten}
\end{equation}
followed by symmetrization of the AO-pair index,
\begin{equation}
P_{M,\mu\nu}
=
\frac12
\left(
\widetilde B_{M,\mu\nu}
+
\widetilde B_{M,\nu\mu}
\right).
\label{eq:ixc_projector}
\end{equation}
This blockwise construction differs from conventional global RI/DF whitening. The goal here
is not to compress the Coulomb operator, but to generate physically organized descriptors:
radial near-linear dependence is removed within each atom-centered $(a,\ell)$ block, while
the angular channels remain identifiable for subsequent invariant contractions.

Using the projector $P_{M,\mu\nu}$, the projected coefficients are
\begin{equation}
c^{(\eta)}_M
=
\Re
\sum_{\mu\nu}
P_{M,\mu\nu}
\left[\mat{D}^{(\eta)}\right]_{\mu\nu},
\qquad
\eta\in\{0,x,y,z\}.
\label{eq:ixc_projected_coeffs}
\end{equation}
The coefficient $c^{(0)}_M$ is the auxiliary-basis coefficient of the Coulomb-projected
charge density, while $c^{(x)}_M$, $c^{(y)}_M$, and $c^{(z)}_M$ are the corresponding
coefficients of the spin magnetization. The resulting model is grid-free and depends on the
electronic state only through linear projections of $\mat{D}$.

The projected coefficients are organized into atom- and $\ell$-resolved blocks
$b=(a,\ell)$. Let $M_{\max}$ denote the fixed upper bound on the number of radial shells in
each block. If a block contains fewer than $M_{\max}$ radial shells, its descriptors are
embedded into fixed-size $M_{\max}\times M_{\max}$ matrices by zero padding. If a block
contains more than $M_{\max}$ radial shells, the descriptor configuration is not defined and
$M_{\max}$ must be increased. In the calculations reported here, $M_{\max}=8$ was
sufficient for all systems considered.

Denoting the shellwise coefficients by $c^{(\eta)}_{b,r,m}$, we build two symmetric Gram
matrices,
\begin{equation}
G^{(n)}_{b,rs}
=
\sum_m
c^{(0)}_{b,r,m}c^{(0)}_{b,s,m},
\qquad
G^{(m)}_{b,rs}
=
\sum_m
\sum_{\eta\in\{x,y,z\}}
c^{(\eta)}_{b,r,m}c^{(\eta)}_{b,s,m}.
\label{eq:ixc_gram_matrices}
\end{equation}
The matrix $\mat{G}^{(n)}$ describes charge-channel couplings between radial shells, while
$\mat{G}^{(m)}$ describes spin-channel couplings. Because $\mat{G}^{(m)}$ depends only on
the scalar combination $c_xc_x+c_yc_y+c_zc_z$, it is invariant under global spin rotations.
Likewise, summation over the magnetic index $m$ gives invariance with respect to rotations
inside the $(2\ell+1)$-dimensional angular subspace of a given shell. These invariances help
preserve physically meaningful degeneracies, especially for states of the same spin
multiplicity \cite{doi:10.1021/acs.jctc.5c00714}.

The block feature vector is obtained by taking the upper triangles of the two Gram matrices
and appending three scalar tags,
\begin{equation}
\vct{x}_b
=
\operatorname{vec}_{\triangle}\!\left(\mat{G}^{(n)}_b\right)
\oplus
\operatorname{vec}_{\triangle}\!\left(\mat{G}^{(m)}_b\right)
\oplus
\left(
\frac{Z_a}{100},
\frac{\ell}{10},
\frac{n_b}{M_{\max}}
\right),
\label{eq:ixc_block_feature}
\end{equation}
where $Z_a$ is the nuclear charge of atom $a$ and $n_b$ is the number of radial shells
actually present in block $b$ before zero padding. For $M_{\max}=8$, the block-feature
dimension is
\begin{equation}
d_b
=
2\times \frac{M_{\max}(M_{\max}+1)}{2}+3
=
75 .
\end{equation}

The neural architecture is additive. Each block feature is first mapped to a hidden
representation,
\begin{equation}
\vct{h}_b
=
\phi_{\mathrm{block}}(\vct{x}_b;\vct{\theta}_{\mathrm{block}})
\in \mathbb R^{w},
\label{eq:ixc_block_network}
\end{equation}
where $\phi_{\mathrm{block}}$ is an MLP with $L$ SiLU-activated hidden layers of width $w$.
The hidden vectors belonging to the same atom are then summed,
\begin{equation}
\vct{u}_a
=
\sum_{b\in\mathcal B(a)}
\vct{h}_b,
\label{eq:ixc_atomic_pool}
\end{equation}
where $\mathcal B(a)$ denotes all $(a,\ell)$ blocks centered on atom $a$. A second MLP maps
the atomwise embedding to a scalar atomic energy,
\begin{equation}
\varepsilon_a
=
\phi_{\mathrm{atom}}
\!\left(
\vct{u}_a \oplus \frac{Z_a}{100};
\vct{\theta}_{\mathrm{atom}}
\right),
\qquad
E_{\mathrm{IXC}}(\mat{D};\vct{\theta})
=
\sum_a
\varepsilon_a .
\label{eq:ixc_model_output}
\end{equation}
The atom network $\phi_{\mathrm{atom}}$ also uses $L$ SiLU-activated hidden layers of width
$w$. All reported models use $L=3$.

The atomwise summation makes the model invariant to the ordering of atoms and provides an
explicitly additive decomposition of the learned correction. In compact form, the model is
\begin{equation}
\mat{D}
\longmapsto
\{c_M^{(\eta)}\}
\longmapsto
\{\mat{G}_b^{(n)},\mat{G}_b^{(m)}\}
\longmapsto
\{\vct{x}_b\}
\longmapsto
\{\vct{h}_b\}
\longmapsto
\{\varepsilon_a\}
\longmapsto
E_{\mathrm{IXC}}(\mat{D};\vct{\theta}).
\label{eq:ixc_pipeline}
\end{equation}
The final affine layer of the atomic head is initialized with zero weights and zero bias, so
that
\begin{equation}
E_{\mathrm{IXC}}(\mat{D};\vct{\theta}_0)=0
\end{equation}
at initialization. Therefore, the initial self-consistent calculation is exactly the
Hartree--Fock baseline with $c_{\mathrm{HF}}=1$.

\subsection{Training}

Training uses three classes of objectives: excitation-energy targets, one-electron
self-interaction constraints, and ground-state energy properties from part of MGCDB84 \cite{Mardirossian2016MinnesotaMGCDB84}. The reference excitation energies are obtained from EOM-CCSD with
the cc-pVDZ basis in PySCF \cite{10.1063/1.464746,10.1063/1.1630018}. The
excitation-energy training set contains 49 molecules. The training dataset we use can be found in our Supporting Information. 

Excitation-energy gradients are differentiated through both the SCF fixed point and the
TDA eigenvalue problem. Total-energy penalty terms are evaluated from fully
self-consistent forward SCF calculations, while their parameter gradients are computed
using the fixed-density derivative.

The reported optimization is staged. In the seed run, excitation, self-interaction error (SIE), and MGCDB84
tasks are all active optimization objectives. In all fine-tuning runs, the excitation
tasks are still evaluated and logged, but they do not contribute gradients; only the
SIE and MGCDB84 objectives update the parameters.

For each molecular excitation sample $n$, the self-consistent ground state is first obtained
and the TDA Casida equation is then solved using the response kernel generated by the same
learned energy functional. From the computed excitation energies and spin expectations
$\{\Omega_k^{(n)},\langle S^2\rangle_k^{(n)}\}$, the first singlet and triplet are selected as
\begin{equation}
\widehat\Omega_{S_1}^{(n)}
=
\min
\left\{
\Omega_k^{(n)}:
\left|
\langle S^2\rangle_k^{(n)}
\right|
\le
\tau_S
\right\},
\qquad
\widehat\Omega_{T_1}^{(n)}
=
\min
\left\{
\Omega_k^{(n)}:
\left|
\langle S^2\rangle_k^{(n)}-2
\right|
\le
\tau_T
\right\}.
\label{eq:ixc_state_selection}
\end{equation}
The reported calculations use $\tau_S=\tau_T=0.75$. The computation of $\langle \hat{S}^2 \rangle$ follows ref. \citenum{doi:10.1021/acs.jctc.6c00314}.

The excitation-energy loss is a scaled Huber objective,
\begin{equation}
\mathcal L_{\mathrm{ex}}^{(n)}(\vct{\theta})
=
\frac{w_{\mathrm{ex}}}{2}
\sum_{\xi\in\{S_1,T_1\}}
\rho_{\delta_{\mathrm{ex}}}
\left(
\frac{
\widehat\Omega_{\xi}^{(n)}(\vct{\theta})
-
\Omega_{\xi,\mathrm{ref}}^{(n)}
}
{\sigma_{\mathrm{ex}}}
\right),
\label{eq:ixc_excitation_loss}
\end{equation}
where
\begin{equation}
\rho_{\delta}(x)
=
\begin{cases}
x^2, & |x|\le \delta,\\
2\delta |x|-\delta^2, & |x|>\delta .
\end{cases}
\label{eq:ixc_huber_loss}
\end{equation}
All excitation energies are expressed in Hartree, whereas
$\mathcal L_{\mathrm{ex}}$ is dimensionless. The excitation block uses
$\sigma_{\mathrm{ex}}=10~\mathrm{m}E_{\mathrm h}$, $\delta_{\mathrm{ex}}=2$, and
$w_{\mathrm{ex}}=1$. This block is active in the seed run and is retained only for
monitoring in the fine-tuning runs.

To impose the one-electron self-interaction condition, the training set is augmented with
one-electron ions from H to Ar, namely H, He$^+$, Li$^{2+}$, $\ldots$, Ar$^{17+}$. For any
one-electron density, full Hartree–Fock exchange exactly cancels the Hartree
self-repulsion. Since the baseline used here has $c_{\mathrm{HF}}=1$, the learned correction
should not introduce any additional one-electron interaction.

The one-electron self-interaction loss is
\begin{equation}
\mathcal L_{\mathrm{SIE}}^{(u)}(\vct{\theta})
=
w_{\mathrm{SIE}}
\left(
\frac{
E_u(\vct{\theta})
-
E_{u,\mathrm{HF}}
}
{\sigma_{\mathrm{SIE}}}
\right)^2 .
\label{eq:ixc_sie_loss}
\end{equation}
Here $E_{u,\mathrm{HF}}$ is the unrestricted Hartree–Fock energy of the one-electron ion
computed in the same basis. The reported runs use
$\sigma_{\mathrm{SIE}}=1~\mathrm{m}E_{\mathrm h}$. The seed run uses
$w_{\mathrm{SIE}}=1$, whereas the fine-tuning runs use $w_{\mathrm{SIE}}=0.03$.

The third component consists of MGCDB84 molecular energy-difference constraints. For an
MGCDB84 sample $m$, with molecular terms $r\in\mathcal R_m$ and signed coefficients
$\nu_{mr}$, the residual is
\begin{equation}
\Delta_m(\vct{\theta})
=
\sum_{r\in\mathcal R_m}
\nu_{mr}
E_{mr}(\vct{\theta})
-
\Delta E_{m,\mathrm{ref}} .
\label{eq:ixc_mgc_residual}
\end{equation}
The corresponding loss is
\begin{equation}
\mathcal L_{\mathrm{MGC}}^{(m)}(\vct{\theta})
=
w_{\mathrm{MGC}}\,
b_m\,
\ell_{\mathrm{MGC}}
\left(
\frac{
\Delta_m(\vct{\theta})
}
{\sigma_{\mathrm{MGC}}}
\right).
\label{eq:ixc_mgc_loss}
\end{equation}
The scaled loss $\ell_{\mathrm{MGC}}$ is stage-dependent. The seed run uses the
Huber loss $\ell_{\mathrm{MGC}}=\rho_{\delta_{\mathrm{MGC}}}$ with
$\sigma_{\mathrm{MGC}}=5~\mathrm{m}E_{\mathrm h}$, $\delta_{\mathrm{MGC}}=5$, and
$w_{\mathrm{MGC}}=10^{-4}$. The fine-tuning runs use the squared loss
$\ell_{\mathrm{MGC}}(x)=x^2$ with
$\sigma_{\mathrm{MGC}}=1.5~\mathrm{m}E_{\mathrm h}$ and
$w_{\mathrm{MGC}}=10^{-3}$.

The factor $b_m$ is a balancing weight. If $m$ belongs to balancing bucket $c(m)$,
then
\[
b_m=\frac{N_{\mathrm{MGC}}}{N_cN_{c(m)}} ,
\]
where $N_{\mathrm{MGC}}$ is the total number of retained MGCDB84 samples, $N_c$ is
the number of balancing buckets, and $N_{c(m)}$ is the number of samples in the
bucket containing $m$. The seed run uses group balancing, whereas the fine-tuning
runs use subset balancing. For the seed run, $N_{\mathrm{MGC}}=80$ and group balancing use
$N_c=4$, with group counts of 28, 24, 12, and 16 for noncovalent,
thermochemistry, isomerization, and barrier samples, respectively. For the fine-tuning runs, $N_{\mathrm{MGC}}=80$ and subset balancing use $N_c=15$. The subset counts are 16 for BDE99nonMR, 8 for X40, 6 for
HB15, S22, BHPERI26, and DBH24, 4 for H2O6Bind8, G21IP, G21EA,
AlkIsomer11, EIE22, Styrene45, and CRBH20, and 2 for XB18 and HW6F.

At each epoch, tasks that fail to converge or produce non-finite energies, states, losses, or
gradients are skipped. Let $q\in\{\mathrm{ex},\mathrm{SIE},\mathrm{MGC}\}$ label the
three task blocks, and let $\mathcal T_{q,t}$ be the set of successful tasks in block $q$
at epoch $t$. The set of active objective blocks is stage-dependent:
\[
\mathcal A_{\mathrm{seed}}=\{\mathrm{ex},\mathrm{SIE},\mathrm{MGC}\},
\qquad
\mathcal A_{\mathrm{ft}}=\{\mathrm{SIE},\mathrm{MGC}\}.
\]
The block-averaged objective is
\begin{equation}
\overline{\mathcal L}_t(\vct{\theta})
=
\sum_{q\in\mathcal A_t}
\frac{1}{|\mathcal T_{q,t}|}
\sum_{i\in\mathcal T_{q,t}}
\mathcal L_i^{(q)}(\vct{\theta}),
\qquad
\vct g_t
=
\nabla_{\vct{\theta}}
\overline{\mathcal L}_t(\vct{\theta}_t),
\label{eq:ixc_block_average}
\end{equation}
where $\mathcal A_t$ contains the active objective blocks with at least one successful task.
Thus, losses are averaged within each active block and then summed across blocks; they are
not averaged over all successful tasks as a single pool. In fine-tuning, excitation losses
are logged but excluded from $\mathcal A_t$.

The gradient is clipped by global norm with $G_{\max}=1$, and an Adam update is formed
with a stage-dependent learning rate $\alpha_{\mathrm{Adam}}$. When the MGCDB84 block is active,
the proposed update is projected to remove any first-order component that would increase
the MGCDB84 block loss. If $\vct g_{\mathrm{MGC},t}$ is the MGCDB84 block-mean gradient,
this projection can be written as
\begin{equation}
\Pi_{\mathrm{MGC}}(\vct u)
=
\vct u
-
\max
\left(
0,
\frac{
\langle \vct g_{\mathrm{MGC},t},\vct u\rangle
}
{
\|\vct g_{\mathrm{MGC},t}\|_2^2+\epsilon_{\mathrm{proj}}
}
\right)
\vct g_{\mathrm{MGC},t},
\label{eq:ixc_mgc_projection}
\end{equation}
where $\epsilon_{\mathrm{proj}}=10^{-30}$. 
If the MGCDB84 block is inactive, or if the update is already non-ascending for this block,
$\Pi_{\mathrm{MGC}}$ reduces to the identity. The trial direction is
\begin{equation}
\vct p_t
=
\Pi_{\mathrm{MGC}}
\left[
\operatorname{AdamStep}
\left(
\operatorname{clip}(\vct g_t;G_{\max});
\alpha_{\mathrm{Adam}}
\right)
\right].
\label{eq:ixc_trial_direction}
\end{equation}

A geometric line search is applied along $\vct p_t$. A candidate $\lambda$ is accepted only if the number of failed tasks does not increase and
the block-averaged objective satisfies
\begin{equation}
\overline{\mathcal L}_t(\vct{\theta}_t+\lambda\vct p_t)
\le
(1+\eta_{\mathrm{LS}})
\overline{\mathcal L}_t(\vct{\theta}_t).
\label{eq:ixc_line_search_accept}
\end{equation}
The accepted update is
\begin{equation}
\vct{\theta}_{t+1}
=
\vct{\theta}_t+\lambda_t\vct p_t .
\label{eq:ixc_optimizer}
\end{equation}
The reported training uses the optimistic variant of this line search: the leading candidate
step is applied provisionally and validated on the next pass over the training set; rejected
provisional steps are rolled back and tested with the same backtracking rule. The
line-search tolerance is $\eta_{\mathrm{LS}}=0.02$ in our reported calculations.

Training was carried out in the following staged sequence:
\begin{center}
\begin{tabular}{lllll}
\toprule
stage & initialization & width $w$ & learning rate & retained step \\
\midrule
seed & random, zero head & 128 & $10^{-4}$ & 229 \\
ft1 & seed step 229 & 128 & $3\times10^{-7}$ & 273 \\
ft2 & ft1 step 273 & 128 & $10^{-6}$ & 167 \\
ft3 & ft2 step 167 & 420 & $3\times10^{-7}$ & 45 \\
ft4 & ft3 step 45 & 420 & $3\times10^{-6}$ & 372 \\
ft5 & ft4 step 372 & 728 & $10^{-7}$ & 293 \\
\bottomrule
\end{tabular}
\end{center}
Only the seed run zero-initializes the final output head. All fine-tuning runs load
saved parameters without resetting the output head. When the width is increased, the
loaded parameters are embedded into the wider network so that the previously learned
subnetwork is preserved at initialization while the added channels remain trainable.
The final retained model used for evaluation is the ft5 model at step 293.

\section{Results and Discussion}
\subsection{Training}
\begin{figure}[H]
  \centering
  \begin{subfigure}[t]{0.48\textwidth}
    \centering
    \includegraphics[width=\linewidth]{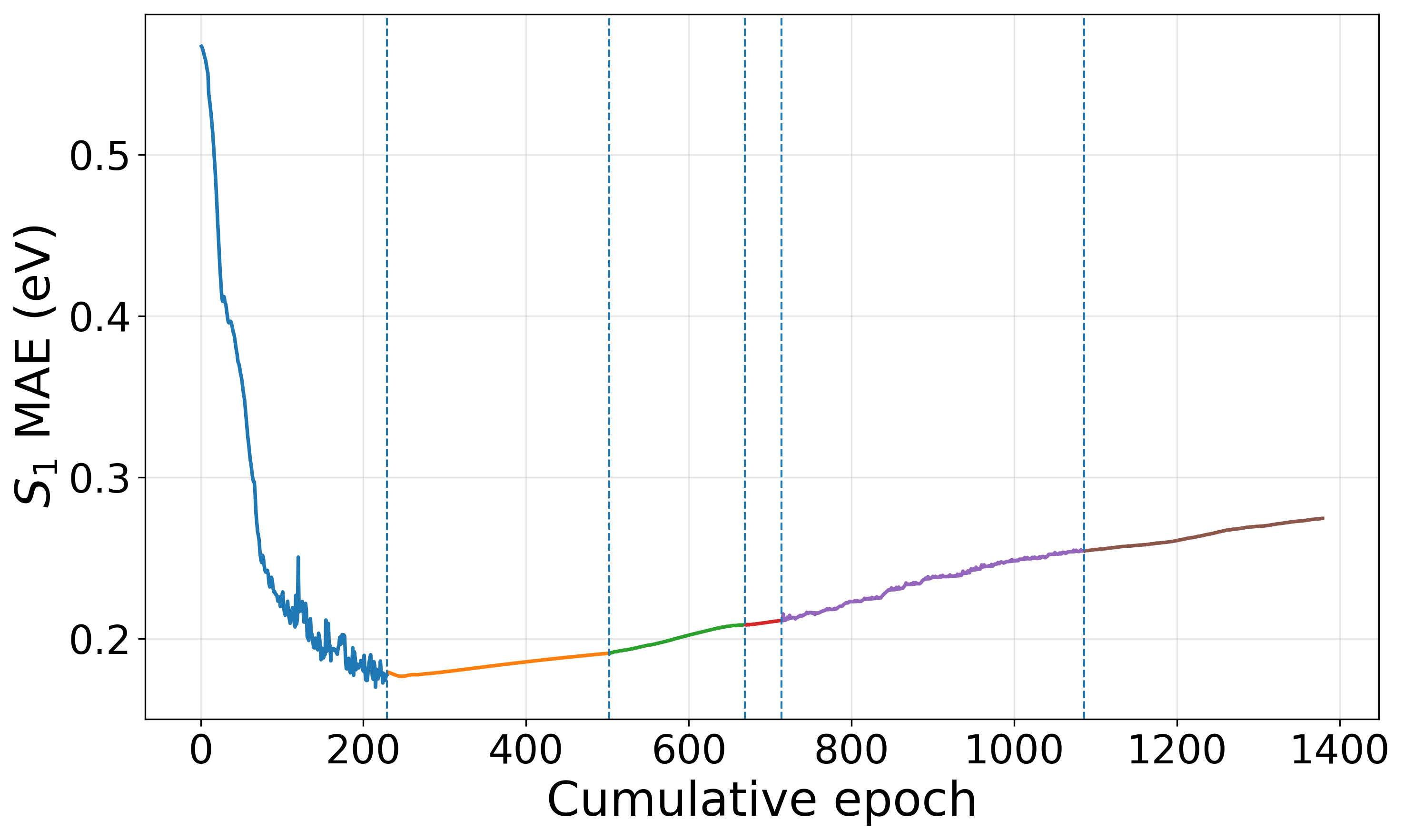}
    \caption{Mean absolute deviation of $\Omega(\mathrm{S_1})$}
    \label{fig:loss_a}
  \end{subfigure}\hfill
  \begin{subfigure}[t]{0.48\textwidth}
    \centering
    \includegraphics[width=\linewidth]{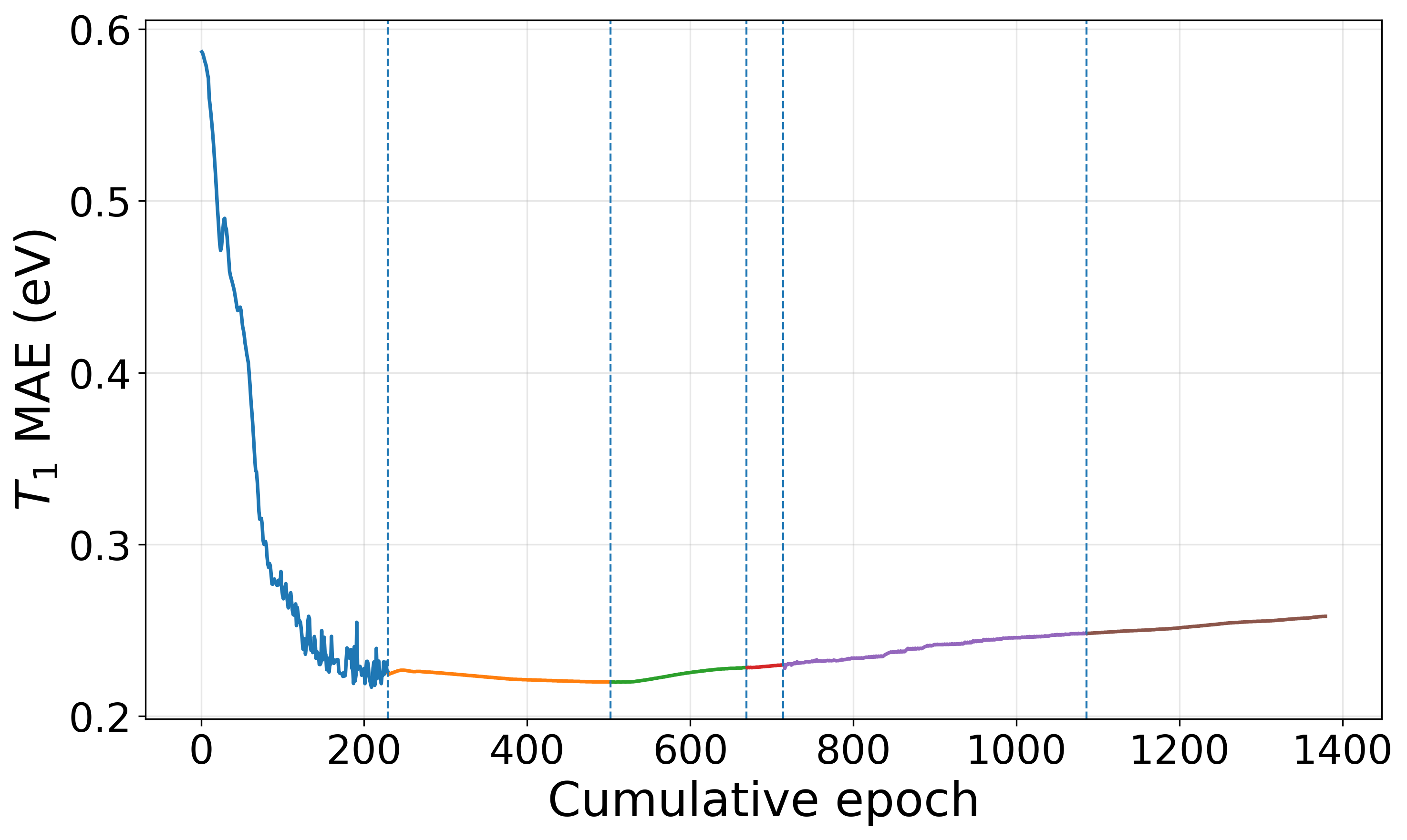}
    \caption{Mean absolute deviation of $\Omega(\mathrm{T_1})$}
    \label{fig:loss_b}
  \end{subfigure}

  \vspace{0.8em}

  \begin{subfigure}[t]{0.48\textwidth}
    \centering
    \includegraphics[width=\linewidth]{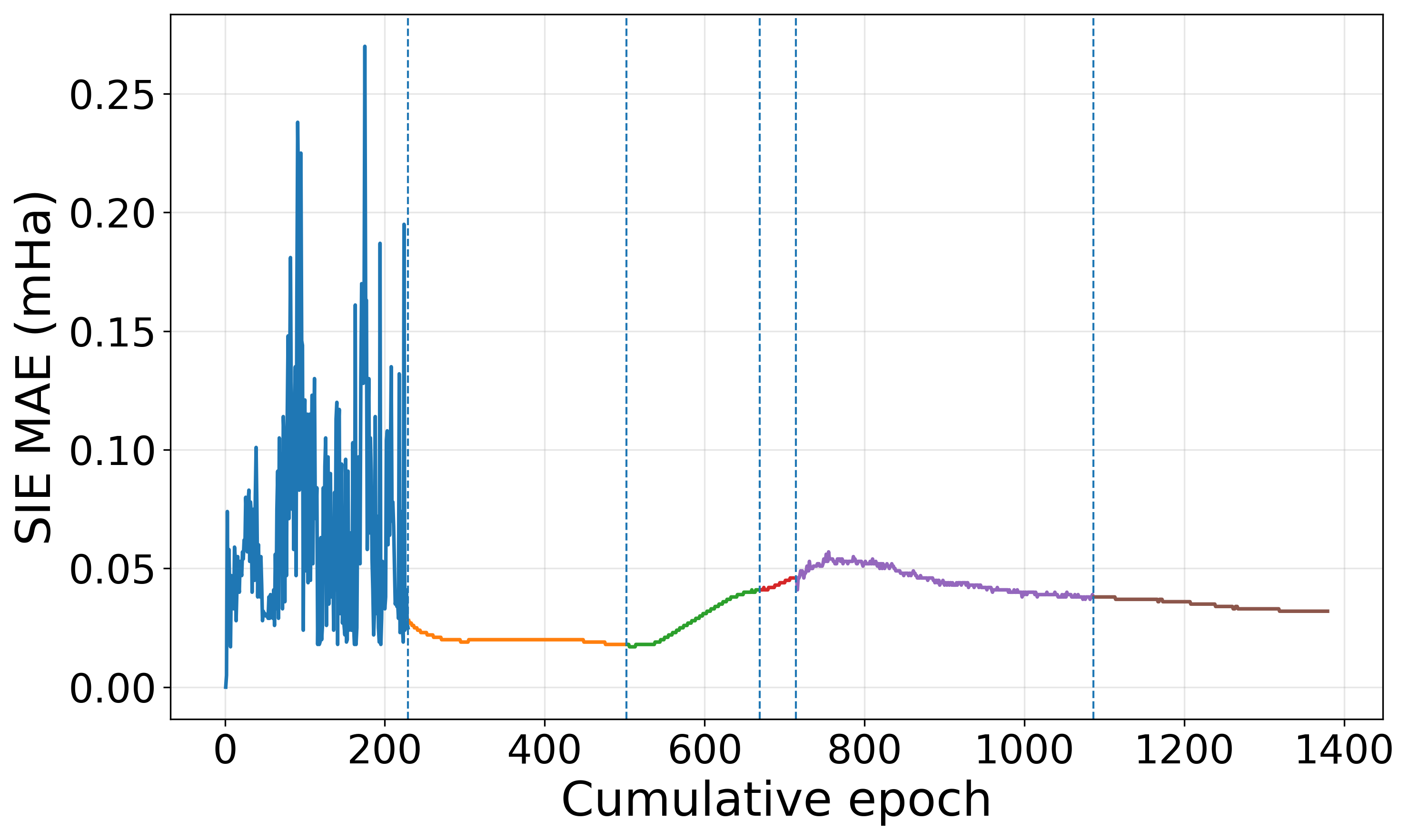}
    \caption{Mean absolute deviation of $\mathrm{SIE}$}
    \label{fig:loss_c}
  \end{subfigure}\hfill
  \begin{subfigure}[t]{0.48\textwidth}
    \centering
    \includegraphics[width=\linewidth]{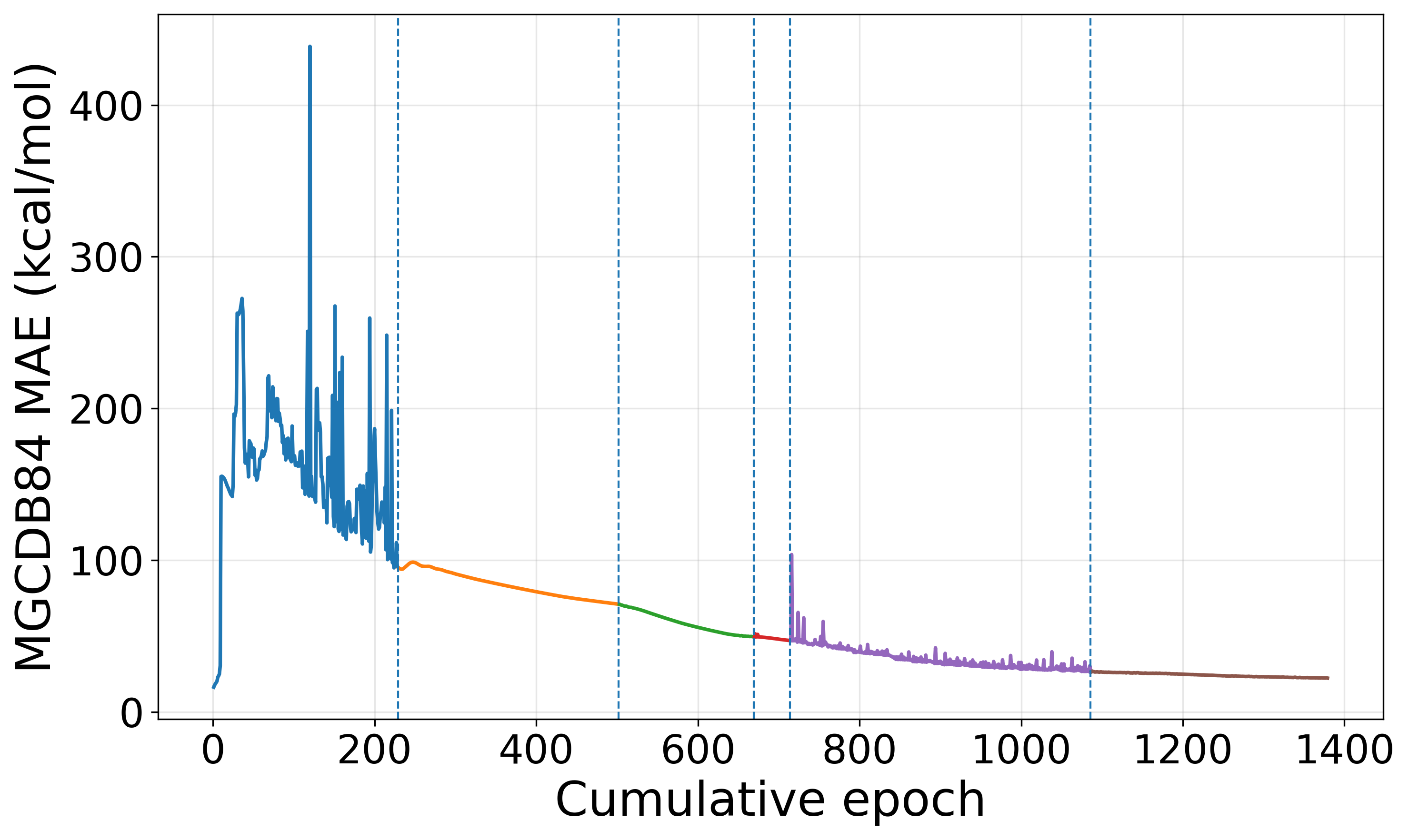}
    \caption{Mean absolute deviation of $\mathrm{MGCDB84}$ core split}
    \label{fig:loss_d}
  \end{subfigure}

  \caption{Training curves for the IXC functional over the full staged optimization}
  \label{fig:loss}
\end{figure}

In Fig.~\ref{fig:loss}, we present the evolution of the mean absolute errors during the full staged training procedure. The four panels report the deviations of $\Omega(\mathrm{S}_{1})$, $\Omega(\mathrm{T}_{1})$, the one-electron self-interaction error, and the MGCDB84 split ground-state energy-difference set, respectively. Calculations are performed with the cc-pVDZ basis. The horizontal axis is the cumulative epoch obtained by concatenating the retained stages of the training trajectory: the seed stage contributes 229 epochs, followed by five fine-tuning stages contributing 273, 167, 45, 372, and 293 epochs, respectively, for a total of 1379 cumulative epochs. The changes in line color, together with the vertical dashed lines, indicate transitions between consecutive training stages.

The seed stage optimizes the excitation-energy, SIE, and MGCDB84 objectives simultaneously. It rapidly reduces the excitation-energy errors, with the $\Omega(\mathrm{S}_{1})$ MAE decreasing from 0.567 eV to 0.178 eV and the $\Omega(\mathrm{T}_{1})$ MAE decreasing from 0.587 eV to 0.226 eV. The later fine-tuning stages optimize only the ground-state SIE and MGCDB84 objectives, while the excitation set is retained as a monitoring set. Consequently, the excitation-energy MAEs increase moderately during fine-tuning, whereas the MGCDB84 MAE decreases substantially from 94.8 kcal/mol at the beginning of fine-tuning to 22.2 kcal/mol in the final retained model. Throughout the retained trajectory, all 147 training tasks at each epoch converge successfully, including 49 excitation tasks, 18 SIE tasks, and 80 MGCDB84 tasks.
The final model selected for evaluation is the retained checkpoint from the last fine-tuning stage. For this model, the MAE of $\Omega(\mathrm{S}_{1})$ is 0.275 eV, the MAE of $\Omega(\mathrm{T}_{1})$ is 0.258 eV, the SIE MAE is 0.032 mHartree, and the MGCDB84 MAE is 22.2 kcal/mol.

Two key observations arise in the training process. First, aggressively accelerating the training on excitation energies deteriorates the accuracy of ground-state predictions. Second, once the model has been optimized for excited states, subsequent fitting of ground states becomes markedly slow and requires substantially more parameters. Consequently, the present study is constrained to relatively small training sets. Attempts to enlarge the dataset or increase model capacity to further reduce the loss render both the required number of epochs and the per-epoch computational cost prohibitive. For these reasons, the development and training of a more general neural-network-based functional are deferred to future work.

\subsection{Excitation Energies}

\begin{table}[H]
\centering
\scriptsize
\resizebox{\textwidth}{!}{%
\begin{tabular}{lrrrrrrrrr}
\toprule
Molecule (State) & TBE & IXC & SPW92 & BLYP & PBE & TPSS & SCAN & B3LYP & PBE0 \\
\midrule
$\mathrm{BeH}(D_1)$     & 2.49 & 2.74 & 2.36 & 2.56 & 2.51 & 2.71 & 2.84 & 2.58 & 2.54 \\
$\mathrm{BH_2}(D_1)$    & 1.18 & 1.44 & 1.10 & 1.36 & 1.33 & 1.59 & 1.79 & 1.34 & 1.33 \\
$\mathrm{CH_3}(D_1)$    & 5.85 & 7.26 & 4.97 & 4.72 & 4.93 & 5.14 & 5.96 & 5.21 & 5.51 \\
$\mathrm{CH_3}(D_2)$    & 6.96 & 7.42 & 5.95 & 5.62 & 5.90 & 6.07 & 6.86 & 6.23 & 6.56 \\
$\mathrm{CH_3}(D_3)$    & 7.18 & 7.66 & 6.34 & 6.71 & 6.81 & 7.25 & 7.65 & 6.87 & 7.04 \\
$\mathrm{CH_3}(D_4)$    & 7.65 & 8.05 & 6.68 & 6.44 & 6.56 & 6.74 & 7.66 & 6.93 & 7.14 \\
$\mathrm{HCl}(S_1)$     & 7.84 & 8.45 & 7.21 & 6.92 & 7.13 & 7.37 & 7.97 & 7.35 & 7.61 \\
$\mathrm{H_2S}(S_1)$    & 6.18 & 6.69 & 5.99 & 5.81 & 5.94 & 6.11 & 6.50 & 5.97 & 6.11 \\
$\mathrm{H_2S}(T_1)$    & 5.81 & 5.94 & 5.59 & 5.34 & 5.42 & 5.57 & 5.81 & 5.47 & 5.54 \\
$\mathrm{H_2S}(T_2)$    & 5.88 & 6.44 & 5.36 & 5.08 & 5.24 & 5.50 & 5.91 & 5.42 & 5.62 \\
$\mathrm{NH_2}(D_1)$    & 2.12 & 2.17 & 2.07 & 2.41 & 2.47 & 2.80 & 3.13 & 2.36 & 2.45 \\
$\mathrm{OH}(D_1)$      & 4.10 & 4.22 & 3.91 & 4.32 & 4.39 & 4.77 & 5.17 & 4.30 & 4.41 \\
$\mathrm{PH_2}(D_1)$    & 2.77 & 2.88 & 2.69 & 2.91 & 2.95 & 3.13 & 3.42 & 2.93 & 3.00 \\
$\mathrm{H_2O}(S_1)$    & 7.62 & 8.40 & 6.58 & 6.27 & 6.42 & 6.60 & 7.29 & 6.93 & 7.19 \\
$\mathrm{H_2O}(T_1)$    & 7.25 & 7.49 & 6.31 & 5.97 & 6.08 & 6.32 & 6.92 & 6.58 & 6.78 \\
$\mathrm{H_2O}(T_2)$    & 9.24 & 9.57 & 7.85 & 7.43 & 7.58 & 7.74 & 8.40 & 8.23 & 8.47 \\
$\mathrm{H_2O}(T_3)$    & 9.54 & 9.74 & 8.30 & 8.01 & 8.13 & 8.39 & 8.92 & 8.64 & 8.87 \\
\midrule
\multicolumn{10}{l}{\textit{Error statistics relative to TBE over 17 matched states (eV)}} \\
MD    & -- &  0.40 & -0.61 & -0.69 & -0.58 & -0.34 &  0.15 & -0.37 & -0.21 \\
MAD   & -- &  0.40 &  0.61 &  0.80 &  0.70 &  0.63 &  0.41 &  0.47 &  0.33 \\
RMSD  & -- &  0.51 &  0.75 &  0.97 &  0.85 &  0.74 &  0.53 &  0.55 &  0.38 \\
\midrule
\multicolumn{10}{l}{\textit{Error statistics excluding the $\mathrm{CH_3}(D_1)$ outlier over 16 matched states (eV)}} \\
MD    & -- &  0.34 & -0.59 & -0.67 & -0.56 & -0.32 &  0.15 & -0.35 & -0.20 \\
MAD   & -- &  0.34 &  0.59 &  0.78 &  0.68 &  0.62 &  0.43 &  0.46 &  0.33 \\
RMSD  & -- &  0.40 &  0.75 &  0.95 &  0.84 &  0.74 &  0.54 &  0.54 &  0.38 \\
\midrule
\multicolumn{10}{l}{\textit{Error statistics for closed-shell molecules over 8 matched states (eV)}} \\
MD    & -- &  0.42 & -0.77 & -1.07 & -0.93 & -0.72 & -0.20 & -0.60 & -0.40 \\
MAD   & -- &  0.42 &  0.77 &  1.07 &  0.93 &  0.72 &  0.32 &  0.60 &  0.40 \\
RMSD  & -- &  0.47 &  0.88 &  1.17 &  1.04 &  0.86 &  0.42 &  0.65 &  0.45 \\
\bottomrule
\end{tabular}%
}
\caption{Excitation energies on a subset of QUEST. All units are eV. TBE means the best estimation. MD is mean deviation. MAD is mean absolute deviation. RMSD is root mean squared deviation.}
\label{tab:exci}
\end{table}

In Table \ref{tab:exci}, we report the performance of our trained IXC functional on excitation energies, which uses Hartree–Fock (HF) as the baseline functional and conducts TDA calculations for a subset of molecules from the QUEST database \cite{doi:10.1021/acs.jctc.5c00975}. Calculations are done with cc-pVDZ. The subset is obtained by restricting the number of non-hydrogen atoms to be less than or equal to one, resulting in a total of 12 molecules. The results are compared with those obtained using the reference SPW92, BLYP, PBE, TPSS, SCAN, B3LYP, PBE0 functionals, which are taken directly from ref. \citenum{doi:10.1021/acs.jctc.2c00160} without further modification. $\mathrm{NH_3}$, Be and CH are not found in this reference, so they are excluded from our table. The corresponding molecular geometries are provided in Appendix A. 

With respect to the full 17-state subset, IXC shows a consistently competitive performance. Although SCAN gives the smallest absolute MD of 0.15 eV and PBE0 gives the lowest MAD and RMSD values of 0.33 and 0.38 eV, respectively, IXC still performs very well, with an MD of 0.40 eV, a MAD of 0.40 eV, and a RMSD of 0.51 eV. In particular, the MAD of IXC is lower than those of SPW92, BLYP, PBE, TPSS, SCAN, and B3LYP, and is only 0.07 eV larger than that of PBE0. Its RMSD is also close to SCAN and B3LYP, and clearly smaller than those of the common semilocal functionals SPW92, BLYP, PBE, and TPSS. After excluding the $\mathrm{CH_3}(D_1)$ outlier, the advantage of IXC becomes even more evident: its MAD and RMSD decrease to 0.34 and 0.40 eV, respectively, which are nearly identical to the best PBE0 values of 0.33 and 0.38 eV, while being clearly better than SCAN, B3LYP, TPSS, PBE, BLYP, and SPW92. For the closed-shell subset, SCAN gives the smallest errors, with MD, MAD, and RMSD values of -0.20, 0.32, and 0.42 eV. Nevertheless, IXC remains highly competitive, giving a MAD of 0.42 eV and a RMSD of 0.47 eV, very close to PBE0 and substantially better than SPW92, BLYP, PBE, TPSS, and B3LYP. Overall, these results indicate that IXC provides robust and balanced excitation energies across both open- and closed-shell systems, and its performance becomes particularly strong once the main outlier is removed.

\subsection{Self-Interaction Error}
\begin{figure}[H]
    \centering
    \begin{subfigure}[t]{0.48\linewidth}
        \centering
        \includegraphics[width=1\linewidth]{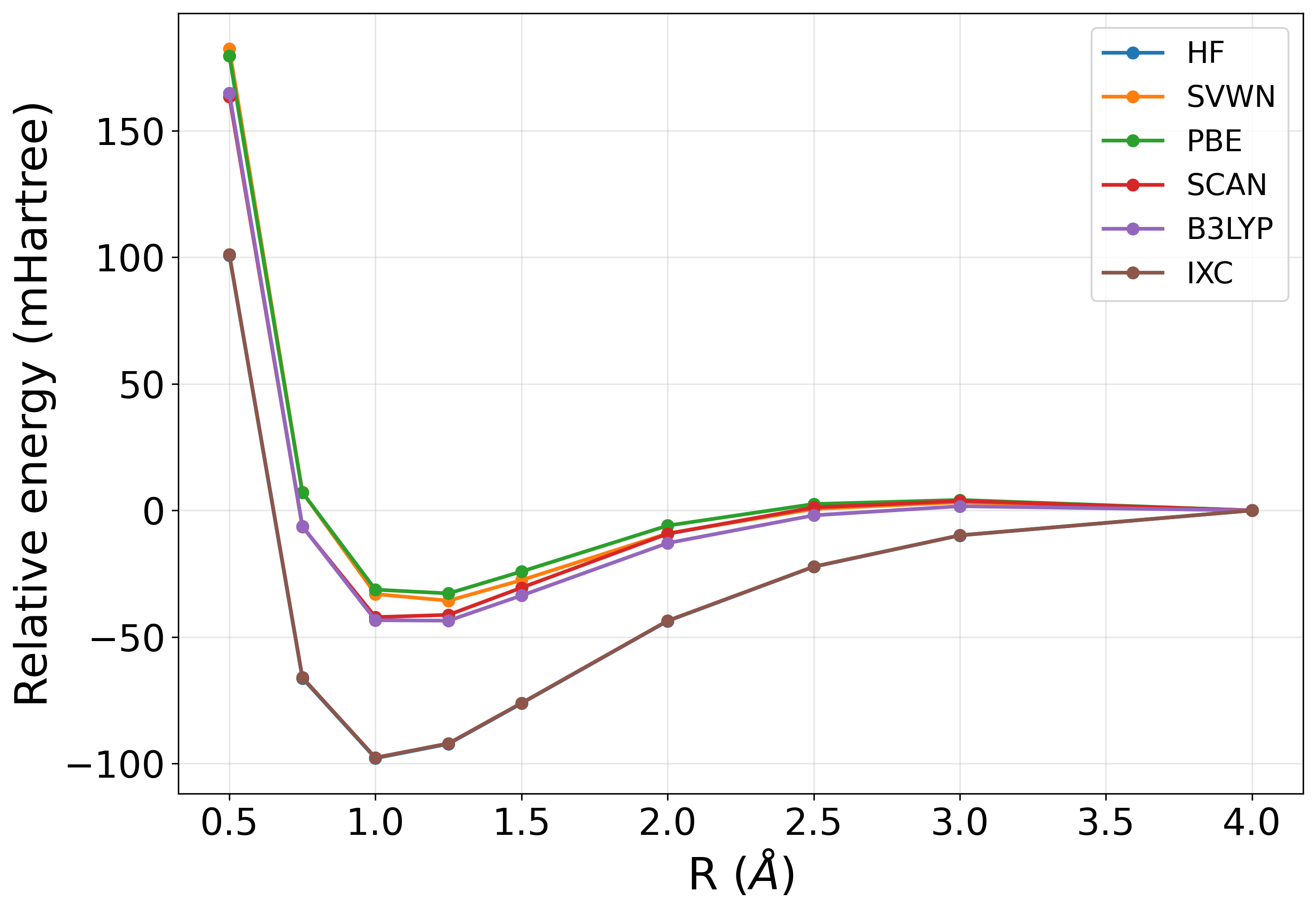}
        \caption{Relative total energy (reference set at $R=4$~\AA).}
    \end{subfigure}\hfill
    \begin{subfigure}[t]{0.48\linewidth}
        \centering
        \includegraphics[width=1\linewidth]{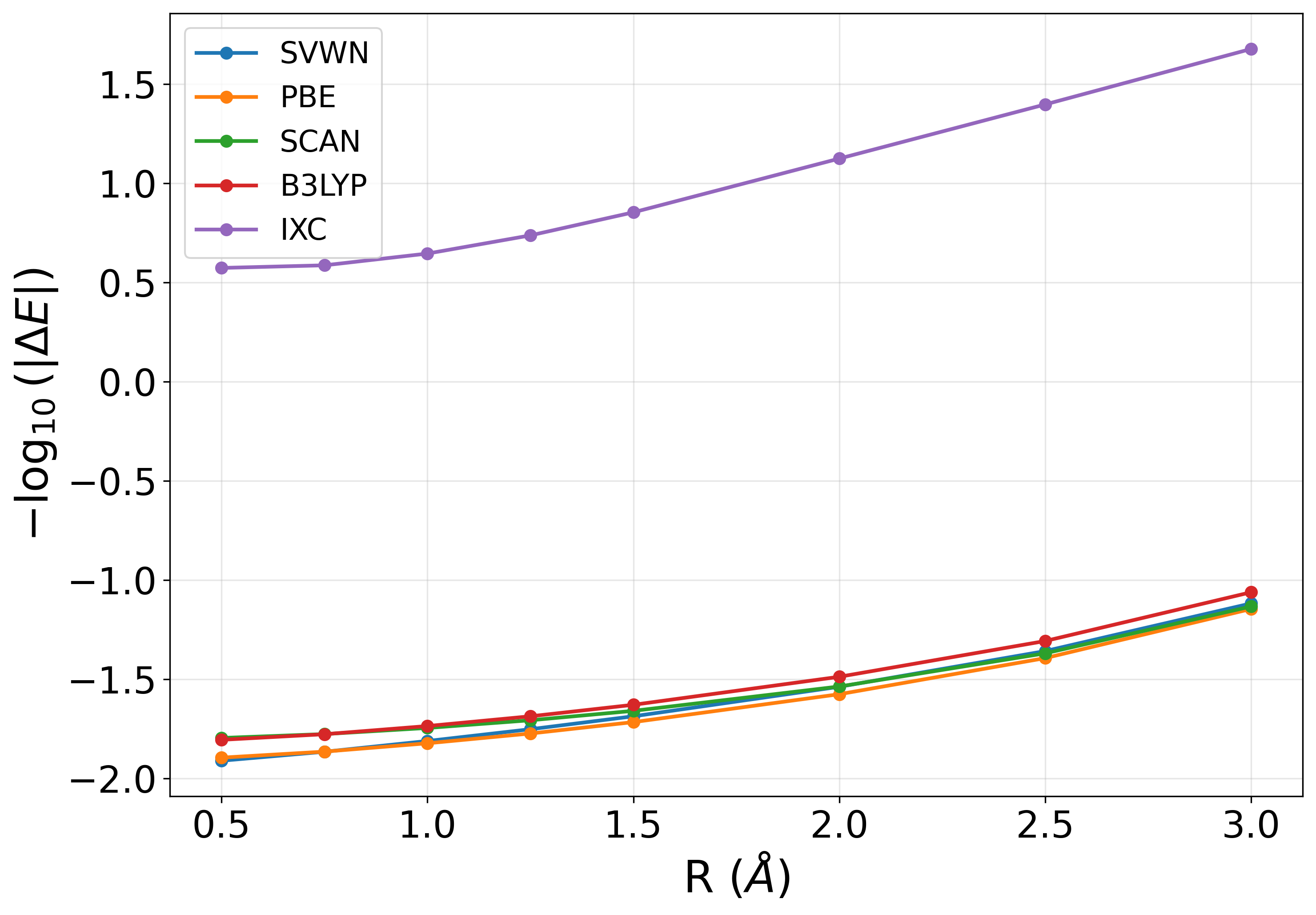}
        \caption{$-\log_{10}(|\Delta E|)$ relative to HF.}
    \end{subfigure}
    \caption{Bond dissociation behavior of $\mathrm{H_2}^+$ calculated with HF, IXC, and traditional exchange--correlation approximations.}
    \label{fig:sie}
\end{figure}

We benchmark density functional theory (DFT) using the IXC functional in combination with the cc-pVDZ basis set for the one-electron dissociation problem of $\mathrm{H_2^+}$. Hartree--Fock (HF) is used as the reference because it is exact for one-electron systems. For comparison, we also evaluate several commonly used exchange–correlation approximations, including the local-density functional SVWN \cite{PhysRev.81.385,LSDA}, the generalized-gradient approximation PBE \cite{PBE}, the meta-GGA functional SCAN \cite{scan}, and the hybrid functional B3LYP \cite{10.1063/1.464913,doi:10.1021/j100096a001}. The dissociation curves are reported in Fig.~\ref{fig:sie}. The relative energies are shifted by taking $E(4.0\,\text{\AA})$ as zero, and the deviations are defined as $\Delta E=E_{\mathrm{method}}^{\mathrm{rel}}-E_{\mathrm{HF}}^{\mathrm{rel}}$. Data for this plot is shown in Appendix B.

The IXC curve closely follows the HF reference over the entire bond-distance range, whereas the conventional xc approximations show large deviations from HF, especially at short and intermediate bond lengths. For example, at $R=1.0\,\text{\AA}$, SVWN, PBE, SCAN, and B3LYP deviate from HF by 64.858, 66.596, 55.736, and 54.524 mHartree, respectively, while IXC deviates by only 0.226 mHartree. Even at the compressed geometry $R=0.50\,\text{\AA}$, the IXC deviation remains only 0.267 mHartree, compared with tens of mHartree for the standard xc approximations. The $-\log_{10}(|\Delta E|)$ plot further highlights this behavior: IXC lies far above the other density functionals because its absolute error relative to HF is orders of magnitude smaller. These results show that IXC substantially suppresses the one-electron self-interaction error and reproduces the HF dissociation profile with high accuracy.

\section{Conclusion}

In summary, we have presented a fully differentiable workflow that enables gradient-based training of a single energy functional using both self-consistent ground-state and adiabatic linear-response (LR) excitation targets. By implementing two-component density functional theory (DFT) and linear-response time-dependent DFT (LR‑TDDFT) in a JAX-based code, and by deriving the corresponding potentials and response kernels via automatic differentiation, the framework enforces analytic consistency among the total energy, self-consistent field (SCF) procedure, and linear-response properties. Owing to the limited size and diversity of the training data, the resulting functional currently lacks broad transferability, although it exhibits clear advantages over traditional functionals in certain test cases. Furthermore, we find that fitting a neural-network-based functional simultaneously for ground and excited states requires a substantially larger number of parameters than fitting it exclusively for ground states, which introduces challenges for model architecture design and for the computational efficiency of the software package. Several promising directions for future work include the development of more mature JAX-based electronic-structure implementations, the construction of more expressive model architectures, the use of more diverse and comprehensive training sets, and the extension of TDDFT beyond the adiabatic approximation.

\begin{acknowledgement}
Xiaoyu Zhang gratefully acknowledges the encouragement provided by Yunlong Xiao. In addition, Xiaoyu Zhang thanks Yixiao Chen for his early-stage yet valuable discussions.
\end{acknowledgement}

\section*{Data and Software Availability}
The data underpinning this study are fully reported in the published article. Our package IQC (v1.0.0) can be obtained from our github release \cite{iqc_user_v100}. Input files and output files are presented in our Supporting Information.

\section*{Supporting Information}

Overview of the supplied reproducibility materials (PDF); README documentation for installation and reproduction; seed-training and fine-tuning scripts; 49-molecule IData training set with optimized geometries and EOM-EE-CCSD S$_1$ and T$_1$ reference excitation energies; H–Ar one-electron-ion self-interaction training cases encoded in the training scripts; 80-sample MGCDB84 core energy-difference training set with molecular geometries and ACCDB source metadata; retained parameter checkpoints and three-rank training logs for the seed run and five successive fine-tuning runs; evaluation scripts and molecular geometries for the excitation-energy calculations; evaluation script for the H$_2^+$ dissociation calculations (ZIP).

\section*{\textbf{Author Information}}

\textbf{Corresponding Author}
\begin{itemize}[leftmargin=16pt, nosep]
    \item Xiaoyu Zhang - College of Chemistry and Molecular Engineering, Peking University, Beijing, 100871, P. R. China.\\
    
    \url{https://orcid.org/0009-0009-4178-3519}; 
    Email: \href{mailto:zhangxiaoyu@stu.pku.edu.cn}{\texttt{zhangxiaoyu@stu.pku.edu.cn}}
\end{itemize}

\bibliography{main}

@article{PhysRevLett.52.997,
  title = {Density-Functional Theory for Time-Dependent Systems},
  author = {Runge, Erich and Gross, E. K. U.},
  journal = {Phys. Rev. Lett.},
  volume = {52},
  issue = {12},
  pages = {997--1000},
  numpages = {0},
  year = {1984},
  month = {Mar},
  publisher = {American Physical Society},
  doi = {10.1103/PhysRevLett.52.997},
  url = {https://link.aps.org/doi/10.1103/PhysRevLett.52.997}
}

@article{10.1063/1.464913,
    author = {Becke, Axel D.},
    title = "{Density‐functional thermochemistry. III. The role of exact exchange}",
    journal = {The Journal of Chemical Physics},
    volume = {98},
    number = {7},
    pages = {5648-5652},
    year = {1993},
    month = {04},
    issn = {0021-9606},
    doi = {10.1063/1.464913},
    url = {https://doi.org/10.1063/1.464913},
    eprint = {https://pubs.aip.org/aip/jcp/article-pdf/98/7/5648/19277469/5648\_1\_online.pdf},
}

@article{YANAI200451,
title = {A new hybrid exchange–correlation functional using the Coulomb-attenuating method (CAM-B3LYP)},
journal = {Chemical Physics Letters},
volume = {393},
number = {1},
pages = {51-57},
year = {2004},
issn = {0009-2614},
doi = {https://doi.org/10.1016/j.cplett.2004.06.011},
url = {https://www.sciencedirect.com/science/article/pii/S0009261404008620},
author = {Takeshi Yanai and David P Tew and Nicholas C Handy}
}

@article{PhysRevResearch.5.013036,
  title = {Noncollinear density functional theory},
  author = {Pu, Zhichen and Li, Hao and Zhang, Ning and Jiang, Hong and Gao, Yiqin and Xiao, Yunlong and Sun, Qiming and Zhang, Yong and Shao, Sihong},
  journal = {Phys. Rev. Res.},
  volume = {5},
  issue = {1},
  pages = {013036},
  numpages = {15},
  year = {2023},
  month = {Jan},
  publisher = {American Physical Society},
  doi = {10.1103/PhysRevResearch.5.013036},
  url = {https://link.aps.org/doi/10.1103/PhysRevResearch.5.013036}
}

@article{RN69,
   author = {Li, Hao and Pu, Zhichen and Sun, Qiming and Gao, Yi Qin and Xiao, Yunlong},
   title = {Noncollinear and Spin-Flip TDDFT in Multicollinear Approach},
   journal = {Journal of Chemical Theory and Computation},
   volume = {19},
   number = {8},
   pages = {2270-2281},
   ISSN = {1549-9618},
   DOI = {10.1021/acs.jctc.3c00059},
   url = {https://doi.org/10.1021/acs.jctc.3c00059},
   year = {2023},
   type = {Journal Article}
}

@article{10.1063/5.0006074,
    author = {Sun, Qiming and Zhang, Xing and Banerjee, Samragni and Bao, Peng and Barbry, Marc and Blunt, Nick S. and Bogdanov, Nikolay A. and Booth, George H. and Chen, Jia and Cui, Zhi-Hao and Eriksen, Janus J. and Gao, Yang and Guo, Sheng and Hermann, Jan and Hermes, Matthew R. and Koh, Kevin and Koval, Peter and Lehtola, Susi and Li, Zhendong and Liu, Junzi and Mardirossian, Narbe and McClain, James D. and Motta, Mario and Mussard, Bastien and Pham, Hung Q. and Pulkin, Artem and Purwanto, Wirawan and Robinson, Paul J. and Ronca, Enrico and Sayfutyarova, Elvira R. and Scheurer, Maximilian and Schurkus, Henry F. and Smith, James E. T. and Sun, Chong and Sun, Shi-Ning and Upadhyay, Shiv and Wagner, Lucas K. and Wang, Xiao and White, Alec and Whitfield, James Daniel and Williamson, Mark J. and Wouters, Sebastian and Yang, Jun and Yu, Jason M. and Zhu, Tianyu and Berkelbach, Timothy C. and Sharma, Sandeep and Sokolov, Alexander Yu. and Chan, Garnet Kin-Lic},
    title = "{Recent developments in the PySCF program package}",
    journal = {The Journal of Chemical Physics},
    volume = {153},
    number = {2},
    pages = {024109},
    year = {2020},
    month = {07},
    issn = {0021-9606},
    doi = {10.1063/5.0006074},
    url = {https://doi.org/10.1063/5.0006074},
    eprint = {https://pubs.aip.org/aip/jcp/article-pdf/doi/10.1063/5.0006074/16722275/024109\_1\_online.pdf},
}

@article{HIRATA1999291,
title = {Time-dependent density functional theory within the Tamm–Dancoff approximation},
journal = {Chemical Physics Letters},
volume = {314},
number = {3},
pages = {291-299},
year = {1999},
issn = {0009-2614},
doi = {https://doi.org/10.1016/S0009-2614(99)01149-5},
url = {https://www.sciencedirect.com/science/article/pii/S0009261499011495},
author = {So Hirata and Martin Head-Gordon}
}

@article{KS-DFT,
  title = {Self-Consistent Equations Including Exchange and Correlation Effects},
  author = {Kohn, W. and Sham, L. J.},
  journal = {Phys. Rev.},
  volume = {140},
  issue = {4A},
  pages = {A1133--A1138},
  numpages = {0},
  year = {1965},
  month = {Nov},
  publisher = {American Physical Society},
  doi = {10.1103/PhysRev.140.A1133},
  url = {https://link.aps.org/doi/10.1103/PhysRev.140.A1133}
}

@article{LSDA,
author = {Vosko, S. H. and Wilk, L. and Nusair, M.},
title = {Accurate spin-dependent electron liquid correlation energies for local spin density calculations: a critical analysis},
journal = {Canadian Journal of Physics},
volume = {58},
number = {8},
pages = {1200-1211},
year = {1980},
doi = {10.1139/p80-159},

URL = { 
    
        https://doi.org/10.1139/p80-159
    
    

},
eprint = { 
    
        https://doi.org/10.1139/p80-159
    
    

}
}

@article{PBE,
  title = {Generalized Gradient Approximation Made Simple},
  author = {Perdew, John P. and Burke, Kieron and Ernzerhof, Matthias},
  journal = {Phys. Rev. Lett.},
  volume = {77},
  issue = {18},
  pages = {3865--3868},
  numpages = {0},
  year = {1996},
  month = {Oct},
  publisher = {American Physical Society},
  doi = {10.1103/PhysRevLett.77.3865},
  url = {https://link.aps.org/doi/10.1103/PhysRevLett.77.3865}
}

@article{PhysRev.81.385,
  title = {A Simplification of the Hartree-Fock Method},
  author = {Slater, J. C.},
  journal = {Phys. Rev.},
  volume = {81},
  issue = {3},
  pages = {385--390},
  numpages = {0},
  year = {1951},
  month = {Feb},
  publisher = {American Physical Society},
  doi = {10.1103/PhysRev.81.385},
  url = {https://link.aps.org/doi/10.1103/PhysRev.81.385}
}

@inbook{tddft,
author = {Casida, Mark E.},
year = {1995},
month = {11},
pages = {155-192},
title = {Time-Dependent Density Functional Response Theory for Molecules},
volume = {1},
isbn = {978-981-02-2442-4},
journal = {Recent Adv. Comput. Chem.},
doi = {10.1142/9789812830586_0005}
}

@article{doi:10.1021/acs.jctc.5c01272,
author = {Zhang, Xiaoyu and Wang, Tai and Gao, Yi Qin and Xiao, Yunlong},
title = {Noncollinear Spin-Flip TDDFT for Potential Energy Surface Crossings: Conical Intersections and Spin Crossings},
journal = {Journal of Chemical Theory and Computation},
volume = {21},
number = {22},
pages = {11550-11561},
year = {2025},
doi = {10.1021/acs.jctc.5c01272},
    note ={PMID: 41208132},

URL = { 
    
        https://doi.org/10.1021/acs.jctc.5c01272
    
    

},
eprint = { 
    
        https://doi.org/10.1021/acs.jctc.5c01272
    
    

}

}

@article{doi:10.1021/acs.jctc.5c00714,
author = {Wang, Tai and Li, Hao and Gao, Yi Qin and Xiao, Yunlong},
title = {Zero Excitation Energy Theorem and the Spin-Flip Kernel},
journal = {Journal of Chemical Theory and Computation},
volume = {21},
number = {14},
pages = {6905-6921},
year = {2025},
doi = {10.1021/acs.jctc.5c00714},
    note ={PMID: 40638888},

URL = { 
    
        https://doi.org/10.1021/acs.jctc.5c00714
    
    

},
eprint = { 
    
        https://doi.org/10.1021/acs.jctc.5c00714
    
    

}

}

@article{doi:10.1021/acs.jctc.5c01305,
author = {Zhang, Xiaoyu and Bao, Taoni},
title = {Operator Formalism for Noncollinear Functionals in the Multicollinear Approach},
journal = {Journal of Chemical Theory and Computation},
volume = {21},
number = {19},
pages = {9620-9630},
year = {2025},
doi = {10.1021/acs.jctc.5c01305},
    note ={PMID: 41039659},

URL = { 
    
        https://doi.org/10.1021/acs.jctc.5c01305
    
    

},
eprint = { 
    
        https://doi.org/10.1021/acs.jctc.5c01305
    
    

}

}

@article{PhysRev.136.B864,
  title = {Inhomogeneous Electron Gas},
  author = {Hohenberg, P. and Kohn, W.},
  journal = {Phys. Rev.},
  volume = {136},
  issue = {3B},
  pages = {B864--B871},
  numpages = {0},
  year = {1964},
  month = {Nov},
  publisher = {American Physical Society},
  doi = {10.1103/PhysRev.136.B864},
  url = {https://link.aps.org/doi/10.1103/PhysRev.136.B864}
}

@article{10.1063/1.1390175,
    author = {Perdew, John P. and Schmidt, Karla},
    title = {Jacob’s ladder of density functional approximations for the exchange-correlation energy},
    journal = {AIP Conference Proceedings},
    volume = {577},
    number = {1},
    pages = {1-20},
    year = {2001},
    month = {07},
    issn = {0094-243X},
    doi = {10.1063/1.1390175},
    url = {https://doi.org/10.1063/1.1390175},
}

@article{10.1063/1.1904565,
    author = {Perdew, John P. and Ruzsinszky, Adrienn and Tao, Jianmin and Staroverov, Viktor N. and Scuseria, Gustavo E. and Csonka, Gábor I.},
    title = {Prescription for the design and selection of density functional approximations: More constraint satisfaction with fewer fits},
    journal = {The Journal of Chemical Physics},
    volume = {123},
    number = {6},
    pages = {062201},
    year = {2005},
    month = {08},
    issn = {0021-9606},
    doi = {10.1063/1.1904565},
    url = {https://doi.org/10.1063/1.1904565},
}

@article{Mardirossian02102017,
author = {Narbe Mardirossian and Martin Head-Gordon},
title = {Thirty years of density functional theory in computational chemistry: an overview and extensive assessment of 200 density functionals},
journal = {Molecular Physics},
volume = {115},
number = {19},
pages = {2315--2372},
year = {2017},
publisher = {Taylor \& Francis},
doi = {10.1080/00268976.2017.1333644},


URL = { 
    
        https://doi.org/10.1080/00268976.2017.1333644
    
    

},

}

@article{Zhao2008TheMS,
  title={The M06 suite of density functionals for main group thermochemistry, thermochemical kinetics, noncovalent interactions, excited states, and transition elements: two new functionals and systematic testing of four M06-class functionals and 12 other functionals},
  author={Yan Zhao and Donald G. Truhlar},
  journal={Theoretical Chemistry Accounts},
  year={2008},
  volume={120},
  pages={215-241},
  url={https://api.semanticscholar.org/CorpusID:98119881}
}

@article{OKUNO201229,
title = {Tuned CAM-B3LYP functional in the time-dependent density functional theory scheme for excitation energies and properties of diarylethene derivatives},
journal = {Journal of Photochemistry and Photobiology A: Chemistry},
volume = {235},
pages = {29-34},
year = {2012},
issn = {1010-6030},
doi = {https://doi.org/10.1016/j.jphotochem.2012.03.003},
url = {https://www.sciencedirect.com/science/article/pii/S101060301200130X},
author = {Katsuki Okuno and Yasuteru Shigeta and Ryohei Kishi and Hiroshi Miyasaka and Masayoshi Nakano}
}

@article{doi:10.1021/ct2009363,
author = {Kronik, Leeor and Stein, Tamar and Refaely-Abramson, Sivan and Baer, Roi},
title = {Excitation Gaps of Finite-Sized Systems from Optimally Tuned Range-Separated Hybrid Functionals},
journal = {Journal of Chemical Theory and Computation},
volume = {8},
number = {5},
pages = {1515-1531},
year = {2012},
doi = {10.1021/ct2009363},
    note ={PMID: 26593646},

URL = { 
    
        https://doi.org/10.1021/ct2009363
    
    

},
eprint = { 
    
        https://doi.org/10.1021/ct2009363
    
    

}

}

@article{neuralxc,
author = {Dick, Sebastian and Fernandez-Serra, Marivi},
year = {2020},
month = {07},
pages = {3509},
title = {Machine learning accurate exchange and correlation functionals of the electronic density},
volume = {11},
journal = {Nature Communications},
doi = {10.1038/s41467-020-17265-7}
}

@article{dm21,
author = {Kirkpatrick, James},
year = {2021},
month = {12},
pages = {},
title = {Pushing the frontiers of density functionals by solving the fractional electron problem},
volume = {374},
journal = {Science},
doi = {10.1126/science.abj6511}
}

@article{doi:10.1021/acs.jctc.0c00872,
author = {Chen, Yixiao and Zhang, Linfeng and Wang, Han and E, Weinan},
title = {DeePKS: A Comprehensive Data-Driven Approach toward Chemically Accurate Density Functional Theory},
journal = {Journal of Chemical Theory and Computation},
volume = {17},
number = {1},
pages = {170-181},
year = {2021},
doi = {10.1021/acs.jctc.0c00872},
    note ={PMID: 33296197},

URL = { 
    
        https://doi.org/10.1021/acs.jctc.0c00872
    
    

},
eprint = { 
    
        https://doi.org/10.1021/acs.jctc.0c00872
    
    

}

}

@article{PhysRevLett.127.126403,
  title = {Learning the Exchange-Correlation Functional from Nature with Fully Differentiable Density Functional Theory},
  author = {Kasim, M. F. and Vinko, S. M.},
  journal = {Phys. Rev. Lett.},
  volume = {127},
  issue = {12},
  pages = {126403},
  numpages = {7},
  year = {2021},
  month = {Sep},
  publisher = {American Physical Society},
  doi = {10.1103/PhysRevLett.127.126403},
  url = {https://link.aps.org/doi/10.1103/PhysRevLett.127.126403}
}

@article{doi:10.1021/acs.jpclett.9b02838,
author = {Zhou, Yi and Wu, Jiang and Chen, Shuguang and Chen, GuanHua},
title = {Toward the Exact Exchange–Correlation Potential: A Three-Dimensional Convolutional Neural Network Construct},
journal = {The Journal of Physical Chemistry Letters},
volume = {10},
number = {22},
pages = {7264-7269},
year = {2019},
doi = {10.1021/acs.jpclett.9b02838},
    note ={PMID: 31690079},

URL = { 
    
        https://doi.org/10.1021/acs.jpclett.9b02838
    
    

},

}

@article{10.1063/1.1535422,
    author = {Wu, Qin and Yang, Weitao},
    title = {A direct optimization method for calculating density functionals and exchange–correlation potentials from electron densities},
    journal = {The Journal of Chemical Physics},
    volume = {118},
    number = {6},
    pages = {2498-2509},
    year = {2003},
    month = {02},
    issn = {0021-9606},
    doi = {10.1063/1.1535422},
    url = {https://doi.org/10.1063/1.1535422},
}

@inproceedings{jax,
  title={Compiling machine learning programs via high-level tracing},
  author={Roy Frostig and Matthew Johnson and Chris Leary},
  year={2018},
  url={https://api.semanticscholar.org/CorpusID:4625928}
}

@article{doi:10.1021/acscentsci.7b00586,
author = {Tamayo-Mendoza, Teresa and Kreisbeck, Christoph and Lindh, Roland and Aspuru-Guzik, Alán},
title = {Automatic Differentiation in Quantum Chemistry with Applications to Fully Variational Hartree–Fock},
journal = {ACS Central Science},
volume = {4},
number = {5},
pages = {559-566},
year = {2018},
doi = {10.1021/acscentsci.7b00586},
    note ={PMID: 29806002},

URL = { 
    
        https://doi.org/10.1021/acscentsci.7b00586
    
    

},
eprint = { 
    
        https://doi.org/10.1021/acscentsci.7b00586
    
    

}

}

@article{PhysRevLett.126.036401,
  title = {Kohn-Sham Equations as Regularizer: Building Prior Knowledge into Machine-Learned Physics},
  author = {Li, Li and Hoyer, Stephan and Pederson, Ryan and Sun, Ruoxi and Cubuk, Ekin D. and Riley, Patrick and Burke, Kieron},
  journal = {Phys. Rev. Lett.},
  volume = {126},
  issue = {3},
  pages = {036401},
  numpages = {7},
  year = {2021},
  month = {Jan},
  publisher = {American Physical Society},
  doi = {10.1103/PhysRevLett.126.036401},
  url = {https://link.aps.org/doi/10.1103/PhysRevLett.126.036401}
}

@article{10.1063/5.0150587,
    author = {Wu, Jiang and Pun, Sai-Mang and Zheng, Xiao and Chen, GuanHua},
    title = {Construct exchange-correlation functional via machine learning},
    journal = {The Journal of Chemical Physics},
    volume = {159},
    number = {9},
    pages = {090901},
    year = {2023},
    month = {09},
    issn = {0021-9606},
    doi = {10.1063/5.0150587},
    url = {https://doi.org/10.1063/5.0150587},
}

@article{10.1063/5.0076202,
    author = {Kasim, Muhammad F. and Lehtola, Susi and Vinko, Sam M.},
    title = {DQC: A Python program package for differentiable quantum chemistry},
    journal = {The Journal of Chemical Physics},
    volume = {156},
    number = {8},
    pages = {084801},
    year = {2022},
    month = {02},
    issn = {0021-9606},
    doi = {10.1063/5.0076202},
    url = {https://doi.org/10.1063/5.0076202},
}

@article{https://doi.org/10.1002/qua.560160825,
author = {Pople, J. A. and Krishnan, R. and Schlegel, H. B. and Binkley, J. S.},
title = {Derivative studies in hartree-fock and møller-plesset theories},
journal = {International Journal of Quantum Chemistry},
volume = {16},
number = {S13},
pages = {225-241},
doi = {https://doi.org/10.1002/qua.560160825},
url = {https://onlinelibrary.wiley.com/doi/abs/10.1002/qua.560160825},
eprint = {https://onlinelibrary.wiley.com/doi/pdf/10.1002/qua.560160825},
year = {1979}
}

@article{RevModPhys.74.601,
  title = {Electronic excitations: density-functional versus many-body Green's-function approaches},
  author = {Onida, Giovanni and Reining, Lucia and Rubio, Angel},
  journal = {Rev. Mod. Phys.},
  volume = {74},
  issue = {2},
  pages = {601--659},
  numpages = {0},
  year = {2002},
  month = {Jun},
  publisher = {American Physical Society},
  doi = {10.1103/RevModPhys.74.601},
  url = {https://link.aps.org/doi/10.1103/RevModPhys.74.601}
}

@article{PhysRevA.101.050501,
  title = {Machine learning exchange-correlation potential in time-dependent density-functional theory},
  author = {Suzuki, Yasumitsu and Nagai, Ryo and Haruyama, Jun},
  journal = {Phys. Rev. A},
  volume = {101},
  issue = {5},
  pages = {050501},
  numpages = {7},
  year = {2020},
  month = {May},
  publisher = {American Physical Society},
  doi = {10.1103/PhysRevA.101.050501},
  url = {https://link.aps.org/doi/10.1103/PhysRevA.101.050501}
}

@article{10.1063/1.464746,
    author = {Stanton, John F. and Bartlett, Rodney J.},
    title = {The equation of motion coupled‐cluster method. A systematic biorthogonal approach to molecular excitation energies, transition probabilities, and excited state properties},
    journal = {The Journal of Chemical Physics},
    volume = {98},
    number = {9},
    pages = {7029-7039},
    year = {1993},
    month = {05},
    issn = {0021-9606},
    doi = {10.1063/1.464746},
    url = {https://doi.org/10.1063/1.464746},
}

@article{10.1063/1.1630018,
    author = {Levchenko, Sergey V. and Krylov, Anna I.},
    title = {Equation-of-motion spin-flip coupled-cluster model with single and double substitutions: Theory and application to cyclobutadiene},
    journal = {The Journal of Chemical Physics},
    volume = {120},
    number = {1},
    pages = {175-185},
    year = {2004},
    month = {01},
    issn = {0021-9606},
    doi = {10.1063/1.1630018},
    url = {https://doi.org/10.1063/1.1630018},
}

@article{scan,
  title = {Strongly Constrained and Appropriately Normed Semilocal Density Functional},
  author = {Sun, Jianwei and Ruzsinszky, Adrienn and Perdew, John P.},
  journal = {Phys. Rev. Lett.},
  volume = {115},
  issue = {3},
  pages = {036402},
  numpages = {6},
  year = {2015},
  month = {Jul},
  publisher = {American Physical Society},
  doi = {10.1103/PhysRevLett.115.036402},
  url = {https://link.aps.org/doi/10.1103/PhysRevLett.115.036402}
}

@article{doi:10.1021/j100096a001,
author = {Stephens, P. J. and Devlin, F. J. and Chabalowski, C. F. and Frisch, M. J.},
title = {Ab Initio Calculation of Vibrational Absorption and Circular Dichroism Spectra Using Density Functional Force Fields},
journal = {The Journal of Physical Chemistry},
volume = {98},
number = {45},
pages = {11623-11627},
year = {1994},
doi = {10.1021/j100096a001},

URL = { 
    
        https://doi.org/10.1021/j100096a001
    
    

},
eprint = { 
    
        https://doi.org/10.1021/j100096a001
    
    

}

}

@article{doi:10.1021/acs.jctc.5c00975,
author = {Loos, Pierre-Fran{\c{c}}ois and Boggio-Pasqua, Martial and Blondel, Aymeric and Lipparini, Filippo and Jacquemin, Denis},
title = {QUEST Database of Highly-Accurate Excitation Energies},
journal = {Journal of Chemical Theory and Computation},
volume = {21},
number = {16},
pages = {8010-8033},
year = {2025},
doi = {10.1021/acs.jctc.5c00975},
    note ={PMID: 40778852},

URL = { 
    
        https://doi.org/10.1021/acs.jctc.5c00975
    
    

},
eprint = { 
    
        https://doi.org/10.1021/acs.jctc.5c00975
    
    

}

}

@article{doi:10.1021/acs.jctc.2c00160,
author = {Liang, Jiashu and Feng, Xintian and Hait, Diptarka and Head-Gordon, Martin},
title = {Revisiting the Performance of Time-Dependent Density Functional Theory for Electronic Excitations: Assessment of 43 Popular and Recently Developed Functionals from Rungs One to Four},
journal = {Journal of Chemical Theory and Computation},
volume = {18},
number = {6},
pages = {3460-3473},
year = {2022},
doi = {10.1021/acs.jctc.2c00160},
    note ={PMID: 35533317},

URL = { 
    
        https://doi.org/10.1021/acs.jctc.2c00160
    
    

},
eprint = { 
    
        https://doi.org/10.1021/acs.jctc.2c00160
    
    

}

}

@Article{B515623H,
author ="Weigend, Florian",
title  ="Accurate Coulomb-fitting basis sets for H to Rn",
journal  ="Phys. Chem. Chem. Phys.",
year  ="2006",
volume  ="8",
issue  ="9",
pages  ="1057-1065",
publisher  ="The Royal Society of Chemistry",
doi  ="10.1039/B515623H",
url  ="http://dx.doi.org/10.1039/B515623H"}

@article{Mardirossian2016MinnesotaMGCDB84,
  author  = {Mardirossian, Narbe and Head-Gordon, Martin},
  title   = {How Accurate Are the Minnesota Density Functionals for Noncovalent Interactions, Isomerization Energies, Thermochemistry, and Barrier Heights Involving Molecules Composed of Main-Group Elements?},
  journal = {Journal of Chemical Theory and Computation},
  year    = {2016},
  volume  = {12},
  number  = {9},
  pages   = {4303--4325},
  doi     = {10.1021/acs.jctc.6b00637}
}

@article{doi:10.1021/acs.jctc.6c00314,
author = {Zhang, Xiaoyu},
title = {A Unified Formulation for ⟨Ŝ2⟩ in Two-Component TDDFT},
journal = {Journal of Chemical Theory and Computation},
volume = {22},
number = {9},
pages = {4429-4438},
year = {2026},
doi = {10.1021/acs.jctc.6c00314},

}

@misc{iqc_user_v100,
  author       = {Zhang, Xiaoyu},
  title        = {{IQC} v1.0.0 Binary User Distribution},
  year         = {2026},
  month        = jul,
  howpublished = {GitHub release},
  url          = {https://github.com/YuleZhang936/iqc-user/releases/tag/v1.0.0},
  note         = {Version v1.0.0}
}

\section*{Appendix}

\subsection*{Appendix A: Geometries of Molecules in Table. \ref{tab:exci}}

\begin{table}[H]
\centering
\begin{tabular}{lrrrr}
\toprule
Molecule & Atom & $x$ & $y$ & $z$ \\
\midrule
\multicolumn{5}{l}{$\mathrm{BeH}$} \\
 & Be & 0.00000000 & 0.00000000 & 0.13284452 \\
 & H  & 0.00000000 & 0.00000000 & -1.18792348 \\
\midrule
\multicolumn{5}{l}{$\mathrm{BH_2}$} \\
 & B  & 0.00000000 & 0.00000000 & 0.07929680 \\
 & H  & 0.00000000 & 1.06427600 & -0.43311221 \\
 & H  & 0.00000000 & -1.06427600 & -0.43311221 \\
\midrule
\multicolumn{5}{l}{$\mathrm{CH_3}$} \\
 & C  & 0.00000000 & 0.00000000 & 0.00000000 \\
 & H  & 0.00000000 & 0.00000000 & 1.07623800 \\
 & H  & 0.00000000 & 0.93205000 & -0.53811900 \\
 & H  & 0.00000000 & -0.93205000 & -0.53811900 \\
\midrule
\multicolumn{5}{l}{$\mathrm{HCl}$} \\
 & Cl & 0.00000000 & 0.00000000 & -0.01317536 \\
 & H  & 0.00000000 & 0.00000000 & 1.26199843 \\
\midrule
\multicolumn{5}{l}{$\mathrm{H_2S}$} \\
 & S  & 0.00000000 & 0.00000000 & -0.26652056 \\
 & H  & 0.00000000 & 0.96219289 & 0.66259489 \\
 & H  & 0.00000000 & -0.96219289 & 0.66259489 \\
\midrule
\multicolumn{5}{l}{$\mathrm{NH_2}$} \\
 & N  & 0.00000000 & 0.00000000 & 0.04231680 \\
 & H  & 0.00000000 & 0.42445251 & -0.29398220 \\
 & H  & 0.00000000 & -0.42445251 & -0.29398220 \\
\midrule
\multicolumn{5}{l}{$\mathrm{OH}$} \\
 & O  & 0.00000000 & 0.00000000 & -0.05749385 \\
 & H  & 0.00000000 & 0.00000000 & 0.91246915 \\
\midrule
\multicolumn{5}{l}{$\mathrm{PH_2}$} \\
 & P  & 0.00000000 & 0.00000000 & 0.06047247 \\
 & H  & 0.00000000 & 1.01549100 & -0.92925852 \\
 & H  & 0.00000000 & -1.01549100 & -0.92925852 \\
\midrule
\multicolumn{5}{l}{$\mathrm{H_2O}$} \\
 & O  & 0.00000000 & 0.00000000 & -0.06990256 \\
 & H  & 0.00000000 & 0.75753241 & 0.51843495 \\
 & H  & 0.00000000 & -0.75753241 & 0.51843495 \\
\bottomrule
\end{tabular}%

\caption{Geometries (in \AA) used in the excitation calculations.}
\end{table}

\subsection*{Appendix B: Relative Energies in Figure. \ref{fig:sie}. }
\begin{table}[H]
\centering

\setlength{\tabcolsep}{3.5pt}
\begin{tabular}{llrrrrrr}
\toprule
$R$ (\AA) & Quantity & HF & SVWN & PBE & SCAN & B3LYP & IXC \\
\midrule
0.50 & $E_\mathrm{rel}$       & 100.831 & 182.375 & 179.526 & 163.442 & 164.802 & 101.098 \\
     & $\Delta E$            &   0.000 &  81.544 &  78.695 &  62.611 &  63.971 &   0.267 \\
0.75 & $E_\mathrm{rel}$       & -66.267 &   7.215 &   7.102 &  -6.415 &  -6.328 & -66.008 \\
     & $\Delta E$            &   0.000 &  73.482 &  73.369 &  59.852 &  59.939 &   0.259 \\
1.00 & $E_\mathrm{rel}$       & -97.902 & -33.044 & -31.306 & -42.166 & -43.378 & -97.676 \\
     & $\Delta E$            &   0.000 &  64.858 &  66.596 &  55.736 &  54.524 &   0.226 \\
1.25 & $E_\mathrm{rel}$       & -92.235 & -35.669 & -32.787 & -41.260 & -43.555 & -92.052 \\
     & $\Delta E$            &   0.000 &  56.566 &  59.448 &  50.975 &  48.680 &   0.183 \\
1.50 & $E_\mathrm{rel}$       & -76.231 & -27.546 & -24.143 & -30.527 & -33.629 & -76.091 \\
     & $\Delta E$            &   0.000 &  48.685 &  52.088 &  45.704 &  42.602 &   0.140 \\
2.00 & $E_\mathrm{rel}$       & -43.672 &  -9.117 &  -6.000 &  -9.228 & -12.906 & -43.597 \\
     & $\Delta E$            &   0.000 &  34.555 &  37.672 &  34.444 &  30.766 &   0.075 \\
2.50 & $E_\mathrm{rel}$       & -22.254 &   0.576 &   2.520 &   1.180 &  -1.943 & -22.214 \\
     & $\Delta E$            &   0.000 &  22.830 &  24.774 &  23.434 &  20.311 &   0.040 \\
3.00 & $E_\mathrm{rel}$       &  -9.861 &   3.264 &   4.136 &   3.738 &   1.667 &  -9.840 \\
     & $\Delta E$            &   0.000 &  13.125 &  13.997 &  13.599 &  11.528 &   0.021 \\
4.00 & $E_\mathrm{rel}$       &   0.000 &   0.000 &   0.000 &   0.000 &   0.000 &   0.000 \\
     & $\Delta E$            &   0.000 &   0.000 &   0.000 &   0.000 &   0.000 &   0.000 \\
\bottomrule
\end{tabular}
\caption{Relative energies of $\mathrm{H_2^+}$ dissociation with $E(4.0\,\text{\AA})$ as zero, and differences from HF. All values are in mHartree. $E_\mathrm{rel}=E(R)-E(4.0\,\text{\AA})$ and $\Delta E=E_\mathrm{method}^{rel}-E_\mathrm{HF}^{rel}$.}
\label{tab:h2p}
\end{table}

\end{document}